\documentclass[twocolumn,english]{revtex4-1}
\usepackage[T1]{fontenc}
\usepackage[latin9]{inputenc}
\usepackage{rotating}
\usepackage{amsmath}
\usepackage{amssymb}
\usepackage{graphicx}

\makeatletter

\providecommand{\tabularnewline}{\\}

\input{Qcircuit}
\usepackage{mathdots}
\usepackage{subfigure}
\usepackage{graphicx}

\makeatother

\usepackage{babel}
\begin{document}

\title{Quantum gates and architecture for the quantum simulation of the
Fermi-Hubbard model}

\author{Pierre-Luc Dallaire-Demers}

\author{Frank K. Wilhelm}

\affiliation{Theoretical physics, Saarland University, 66123 Saarbrücken, Germany}

\date{\today}
\begin{abstract}
Quantum computers are the ideal platform for quantum simulations.
Given enough coherent operations and qubits, such machines can be
leveraged to simulate strongly correlated materials, where intricate
quantum effects give rise to counter-intuitive macroscopic phenomena
such as high-temperature superconductivity. In this paper, we provide
a gate decomposition and an architecture for a quantum simulator used
to simulate the Fermi-Hubbard model in a hybrid variational quantum-classical
algorithm. We propose a simple planar implementation-independent layout
of qubits that can also be used to simulate more general fermionic
systems. By working through a concrete application, we show the gate
decomposition used to simulate the Hamiltonian of a cluster of the
Fermi-Hubbard model. We briefly analyze the Trotter-Suzuki errors
and estimate the scaling properties of the algorithm for more complex
applications.
\end{abstract}
\maketitle

\section{Introduction\label{sec:Introduction}}

Simulating quantum phenomena with classical computers is often hard.
This observation originated the idea of universal quantum simulators,
or quantum computers \citep{Feynman82}. Since then, technology has
advanced to the point where small collections of interacting quantum
bits (qubits) can be fabricated, characterized and controlled to find
the ground state energy of simple molecules \citep{OMalley15} in
quantum chemistry. Scaling to a few tens or hundreds of highly coherent
qubits will open new ways to study classes of important but classically
intractable problems. The prototypical non-integrable system where
long-range entanglement and short-range fluctuations makes classical
simulation prohibitive is the two-dimensional Fermi-Hubbard model,
where electrons can hop on a bipartite lattice with local Coulomb
interaction \citep{Hubbard63}. The Fermi-Hubbard model can be used
to explain phenomena arising in Mott insulators and cuprate superconductors
\citep{Anderson87}. As in the simulation of quantum chemistry \citep{Kassal11,Peruzzo14,OMalley15,Babbush16},
the simulation of strongly correlated materials can also be improved
by hybrid quantum-classical solver \citep{LasHeras13,Lamata14,Barends15,LasHeras15,Bauer15,Salathe15}
even if the number of interacting particles is in principle macroscopic. 

To study phase transitions occuring in condensed matter systems, single-particle
correlation functions containing the information of the dynamics of
the excited states have to be computed. Correlated lattices can be
approximated in variational algorithms by constraining the space of
possible self-energies to that of a lattice of finite clusters \citep{Potthoff03}.
The solutions can be refined systematically by increasing the size
of the clusters, however the classical memory required to represent
state vectors in the clusters Hilbert space increases exponentially
with the number of simulated electronic orbitals in a cluster. We
showed in earlier work \citep{DDW16} how to extend the range of applicability
of variational classical cluster methods by leveraging small quantum
computers. The quantum algorithm uses black-box time evolutions\citep{Thompson13}
without making any assumptions on the architecture of the underlying
quantum computer. This paper is meant to extend the quantum algorithm
and to present a natural architecture and gate decomposition as an
example to a general-purpose quantum simulator for dynamical cluster
methods. Such a device could significantly improve our capabilities
to investigate and simulate the macroscopic properties of correlated
systems of electrons.

Here we present four main results. First, a practical physical layout
of the simulator can be made with two parallel chains of qubits with
nearest-neighbor interactions and a control/probe qubit connected
to all elements of the chains. The layout is fabrication-friendly
as it has no crossing interaction lines, yet it can simulate Gibbs
states of a lattice of arbitrary dimensionality. Second, there is
a limited number of three-qubit gates that need to be tuned and benchmarked
prior to a simulation, these gates are called ``conditional imaginary
swap'' ($\mathrm{c-\pm iSWAP}$ or $\mathrm{iFredkin}$) with positive
and negative varieties. Third, the toughest terms of a cluster Hamiltonian
can be decomposed in a number of gates which is subquadratic in the
size of the cluster. Finally, a numerical example is used to show
that the Trotter-Suzuki approximations can reach arbitrary precision
when non-commuting terms in the cluster Hamiltonian are propagated
in time.

Specifically, the paper is structured in the following way. In section
\ref{sec:FHMandGates}, the Fermi-Hubbard is briefly introduced. In
subsection \ref{sub:QuantumComputer}, the core elements of the quantum
solver are reviewed and an architecture is proprosed for a quantum
simulator. In section \ref{sec:TimeEvolution} the gate decomposition
of the time evolution of the cluster is given through the example
of a $2\times2$ Fermi-Hubbard cluster. The Jordan-Wigner transformation
used is shown in subsection \ref{sub:HamiltonianCluster} and subsection
\ref{sub:MeasuringCorrelationFunction} introduces the notation used
in the procedure to measure the correlation function and more notation
concerning the mapping of qubits to spin orbitals. The explicit gate
decomposition of important terms of the Fermi-Hubbard model are given
in subsection \ref{sub:GateDecomposition}. A short analysis of Trotter-Suzuki
errors is done in subsection \ref{sub:TrotterSuzuki}. Finally, the
scaling properties of the quantum ressources involved in scaling the
algorithm are analysed in section \ref{sec:Scaling}.

\section{Solving the Fermi-Hubbard model on a quantum computer\label{sec:FHMandGates}}

The model describes a simple electronic band in a periodic square
lattice where electrons are free to hop between orbitals (or sites)
with kinetic energy $t$ and interact via a simple two-body Coulomb
term $U$. The standard form of the Fermi-Hubbard Hamiltonian is given
by

\begin{equation}
\mathcal{H}=-t\sum_{\left\langle i,j\right\rangle ,\sigma}c_{i\sigma}^{\dagger}c_{j\sigma}+U\sum_{i}n_{i\uparrow}n_{i\downarrow}-\mu\sum_{i,\sigma}n_{i\sigma},\label{eq:FermiHubbardHamiltonian}
\end{equation}
where $\mu$ is the chemical potential that controls the occupation
of the band. The $c_{i\sigma}$($c_{i\sigma}^{\dagger}$) are the
fermionic annihilation (creation) operators and the number operators
are $n_{i\sigma}=c_{i\sigma}^{\dagger}c_{i\sigma}$. Note that in
the rest of this document, units are chosen such that $\hbar=1$ and
$k_{B}=1$. The hopping energy $t=1$ is assumed to be the reference
energy and inverse time unit. The model is analytically solvable in
the tight-binding limit $\frac{U}{t}\rightarrow0$ and the atomic
limit $\frac{t}{U}\rightarrow0$. For a finite $\frac{U}{t}$, there
is competition from different orders (antiferromagnetism, superconductivity)
and no general solution is known for more than one dimension \citep{Lieb03}.
Many numerical methods have been developped to compute the thermodynamic
properties of the Fermi-Hubbard model \citep{Senechal05,Tremblay06}.
Dynamical mean field methods, unified under the broader self-energy
functional theory, can asymptotically approach solutions of the model
by simulating the dynamics of increasingly larger clusters that contain
the information of the quantum fluctuations of the system. However,
simulating those clusters on a classical computer is a task that requires
an exponential amount of computing ressources as the cluster size
is increased. A general introduction to the classical cluster methods
and self-energy functional theory can be found in \citep{Potthoff03,Senechal08}.
In \citep{DDW16}, we showed how the important information of the
clusters could be extracted from a quantum computer. In the next part,
we explain how an architecture can be chosen for a quantum simulator
such that the time evolution of any cluster Hamiltonian becomes very
natural.

\subsection{The layout of qubits\label{sub:QuantumComputer}}

\begin{figure}
\begin{centering}
\begin{minipage}[t]{1\columnwidth}%
\begin{center}
\subfigure[]{\label{fig:FullCircuit}\includegraphics[width=3.375in]{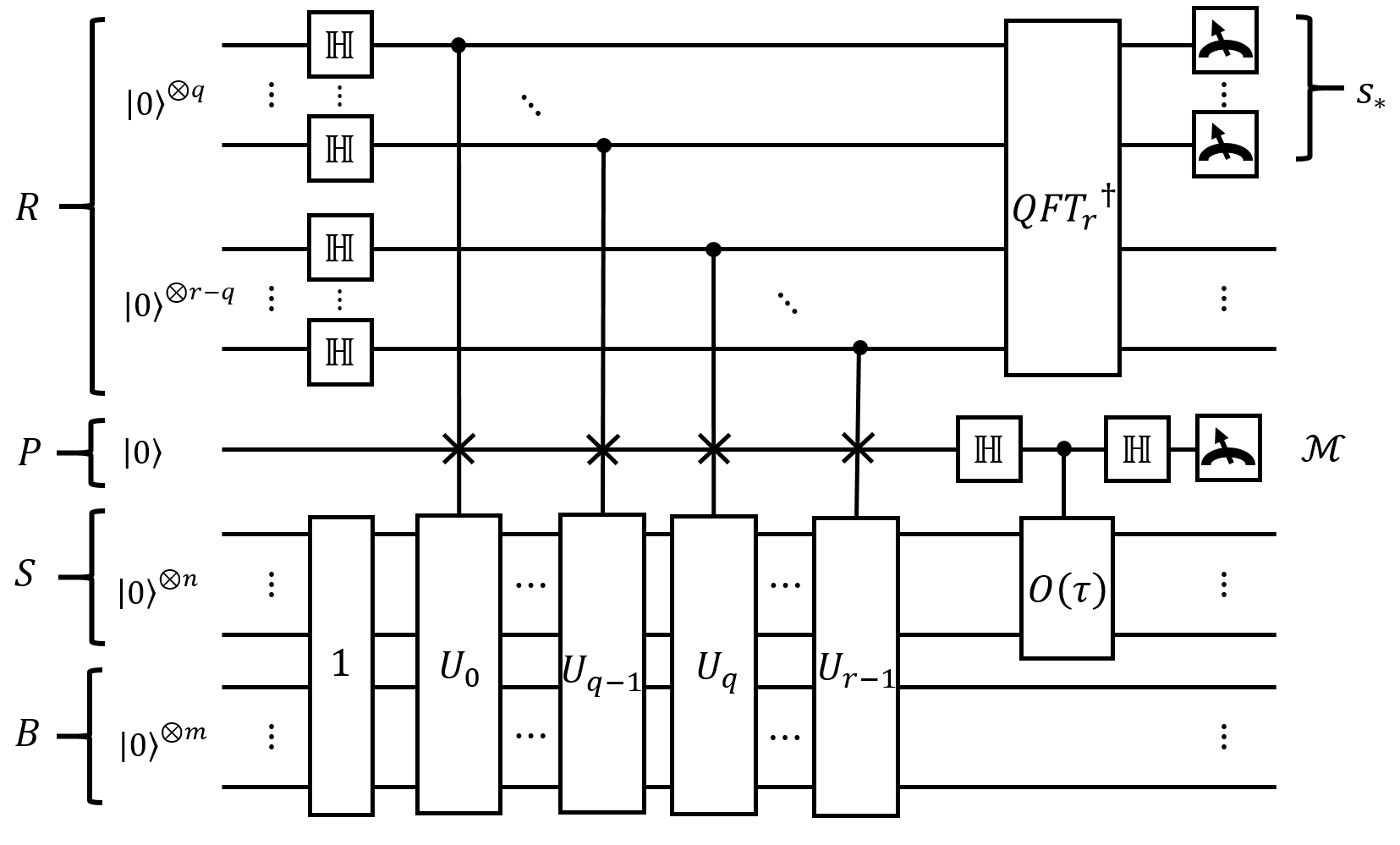}}
\par\end{center}%
\end{minipage}
\par\end{centering}

\smallskip{}
\begin{minipage}[t]{1\columnwidth}%
\subfigure[]{\label{fig:InteractionThroughP}\Qcircuit @C=1em @R=.7em @!R { \lstick{R} & \ctrl{2} & \qw & & & \lstick{R}&\qswap & \qw & \qswap & \qw\\ \lstick{P} & \qswap & \qw & \push{\rule{.3em}{0em}\equiv\rule{.3em}{0em}} & & \lstick{P}& \qswap \qwx & \ctrl{1} & \qswap \qwx & \qw \\ \lstick{Q} & \gate{U} & \qw & & & \lstick{Q}&\qw & \gate{U} & \qw & \qw  \\ & & & & & & & & &}}%
\end{minipage}

\caption{In \subref{fig:FullCircuit}, the circuit used to simulate the time-dependent
correlation function of the cluster Hamiltonian (\ref{eq:HamiltonianOneCluster})
is shown. The first part meant to generate a Gibbs state is taken
from \citep{Riera12}. Register $R$ is used in the modified phase
estimation scheme to prepare a rectangular state between the bath
and the system contained in register $Q$. When the bath is traced
out the system channel is left in a Gibbs state from which the different
correlation functions can be read from the one-qubit register $P$.
The size of register $Q$ depends on the number of orbitals in the
simulated cluster (typically $n=2L_{c}$) and the bath size (which
should be some constant factor larger than the system register). Register
$R$ is used as a digital component and $q$ should therefore be the
size required for the desired floating point accuracy on reading $s_{*}$.
Note that the numbers in the controlled gates of register $R$ denote
the index of the qubit which is acting as the control. Figure \subref{fig:InteractionThroughP}
shows how the interaction through register $P$ is done. In total,
$2q$ $\mathrm{SWAP}$ gates are required. Alternatively, only one
swap per step can be used if the initial Hadamard gates from figure
\ref{fig:FullCircuit} are done directly on $P$.\label{fig:AlgoCircuit}}
\end{figure}

\begin{figure}
\begin{centering}
\includegraphics[width=3.375in]{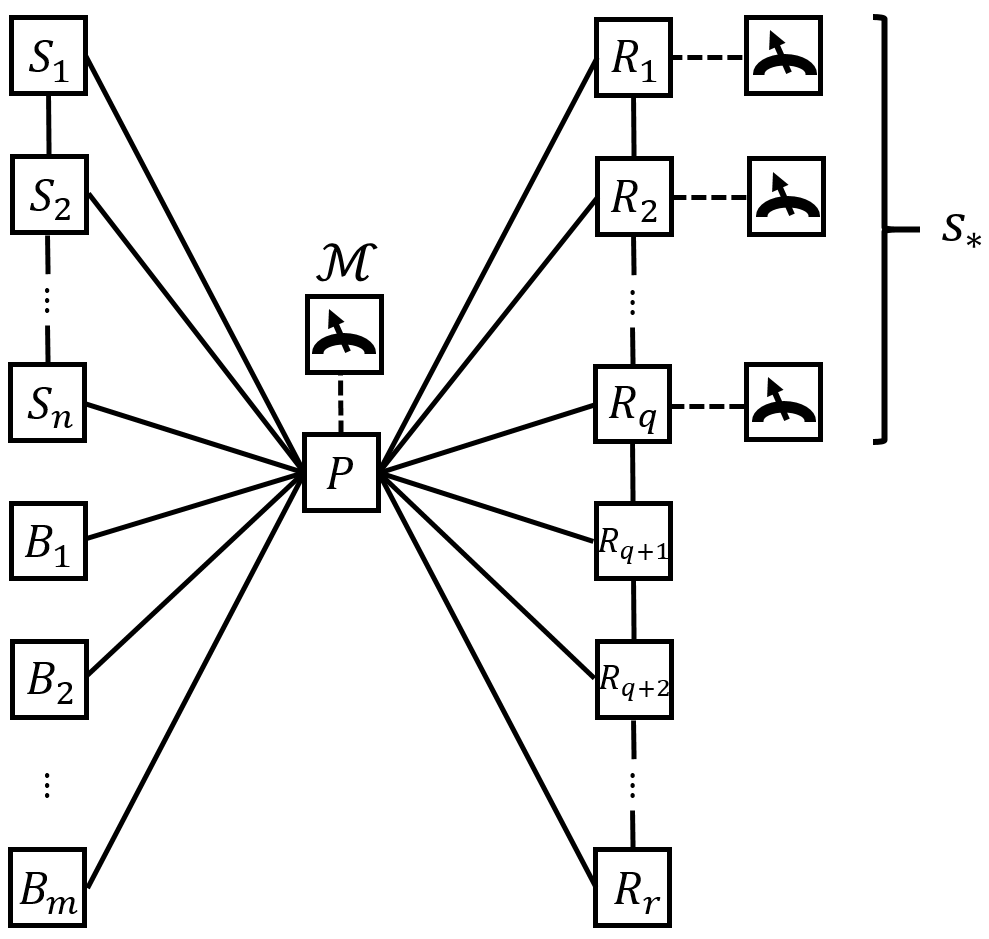}
\par\end{centering}

\caption{Proposed layout of physical qubits with no crossing interaction line.
Boxes represent physical qubits in different labelled registers. Arbitrary
single qubit gates are assumed to be implementable on every qubit.
Solid lines are tunable exchange interactions ($\sigma_{x}\otimes\sigma_{x}+\sigma_{y}\otimes\sigma_{y}$).
Early numerical work also suggests using tunable dispersive interactions
($\sigma_{z}\otimes\sigma_{z}$) for the $S-P$ and $B-P$ connections
to implement the required conditional two-qubit gates more efficiently.
The interactions between the qubits in registers $S$ (or $B$) and
the qubit in $P$ are used to implement conditional $\mathrm{\pm iSWAP}$s
and controlled single-qubit gates. The interactions between the qubits
in register $R$ and the one in register $P$ are only used to implement
$\mathrm{SWAP}$ gates. The interactions between the qubits in $R$
are used to implement $\mathrm{QFT^{\dagger}}$ on this register.
Dashed lines are linked to qubits that are measured in the computational
basis at the end of the protocol. There are only a very limited number
of gates to benchmark and tune. The size the register $R$ depends
on the desired precision and accuracy of the Gibbs state preparation
(floating point accuracy should roughly correspond to the quantum
supremacy crossover for this register). The size of register $S$
should be at least as large as the number of spin orbitals in the
simulated cluster Hamiltonian and the size of register $B$ is equal
to the size of register $S$ such that it can absorb the excess entropy
of the Gibbs state preparation.\label{fig:Layout}}
\end{figure}

We introduced an hybrid quantum-classical solver in \citep{DDW16}
to show how some parts of quantum cluster methods can be improved
by executing them on universal quantum computers. We refer the readers
to our previous work for discussion and details of the various parameters.
In the present paper we show that there is a simple physical layout
of qubits which implements naturally the quantum circuit of figure
\ref{fig:FullCircuit}. The circuit is used to prepare a Gibbs state
of a cluster of the Fermi-Hubbard model in register $S$ and output
the single-particle correlation functions in register $P$ (the operator
$O\left(\tau\right)$ is detailed in section \ref{sub:MeasuringCorrelationFunction}).
In principle the same type of circuit can simulate many other physical
models, the Fermi-Hubbard model is used as an example that encapsulates
the essence of strongly correlated systems. Since each register performs
a definite task in the algorithm, the qubit layout can also be divided
into modules. A subtle but important difference to \citep{DDW16}
consists in controlling the bath ($B$) + system ($S$) registers
through qubit $P$. This significantly reduces the number of elements
that have to be controlled on the quantum simulator chip. Since the
extraction of the correlation functions is done by measuring the probability
of $\mathcal{M}=1$ and given that $S$ is in general in a mixed state,
there is no clear advantage to using more than one qubit in register
$P$. It can therefore be used to mediate the operations between register
$R$ and $S+B$ in the Gibbs state preparation protocol (see figure
\ref{fig:InteractionThroughP}). The suggested physical layout of
qubits is shown in figure \ref{fig:Layout}, the qubits of $R$ and
$S+B$ are aligned as parallel chains with nearest-neighbor interactions
and all conditional operations from $R$ are mediated through qubit
$P$. An important feature of the proposed physical layout is the
absence of overlapping interaction lines. Compared to a general purpose
quantum computer, a dedicated quantum circuit has a much smaller set
of gates that have to be tuned and benchmarked to solve a class of
problems. Register $R$ needs only to support single qubits Hadamard
gates and the operations required for an inverse quantum Fourier transform
($\mathrm{QFT^{\dagger}}$), only $q$ qubits are measured to determine
the effective temperature $\beta$ of the Gibbs state prepared (depending
on the output $s_{*}$, see \citep{Riera12} for details). The operations
between $P$ and $B$ can all be reduced to controlled single qubits
phase rotations as the bath is assumed to consists of independent
spins. The operations between $P$ and $S$ require a more detailed
analysis.

First, a one dimensional chain of qubits with local controls and nearest
neighbor exchange interaction is sufficient to implement the simulation
of a higher dimensional cluster of a correlated electrons system.
The exchange interaction can be used to generate the $\mathrm{iSWAP}$
gate which can be used to implement any Pauli string arising from
the Jordan-Wigner form of given fermionic cluster Hamiltonians \citep{Kaicher16}.
Two dimensional clusters of the Fermi-Hubbard model can be simulated
efficiently with a number of gates which scales sub-quadratically
with the number of orbitals. Finally, using a Trotter-Suzuki decomposition,
the time evolution can be implemented accurately and with a better
scaling than typical ``hard'' molecules \citep{Poulin15}.

\subsection{Jordan-Wigner transformation\label{sub:JordanWigneTransformation}}

Qubits in quantum computer are distinguishable objects, while electrons
are not. In order to map the fermionic creation and annihilation operators
of the Hamiltonian to the computational basis, a Jordan-Wigner transformation
\citep{Jordan28} can be used. If there are $n=2L_{c}$ electrons,
then the Jordan-Wigner transformed creation operators are given by

\begin{equation}
\begin{array}{ccl}
c_{i\uparrow}^{\dagger} & = & \mathbb{I}^{\otimes2\left(L_{c}-i\right)+1}\otimes\sigma_{+}\otimes\sigma_{z}^{\otimes2\left(i-1\right)}\\
\\
c_{i\downarrow}^{\dagger} & = & \mathbb{I}^{\otimes2\left(L_{c}-i\right)}\otimes\sigma_{+}\otimes\sigma_{z}^{\otimes2i-1}
\end{array}.\label{eq:JordanWignerMap}
\end{equation}
This ensures that the fermionic anticommuation relation $\left\{ c_{i\sigma},c_{j\sigma'}^{\dagger}\right\} =\delta_{ij}\delta_{\sigma\sigma'}$
and $\left\{ c_{i\sigma},c_{j\sigma'}\right\} =\left\{ c_{i\sigma}^{\dagger},c_{j\sigma'}^{\dagger}\right\} =0$
are enforced. In this notation,

\begin{equation}
\mathbb{\sigma}^{\otimes k}\equiv\begin{cases}
1 & k=0\\
\sigma & k=1\\
\sigma\otimes\mathbb{\sigma}^{\otimes k-1} & k>1
\end{cases},\label{eq:TensorExponentNotation}
\end{equation}
also $\sigma_{+}=\frac{\left(\sigma_{x}+i\sigma_{y}\right)}{2}$,
$\sigma_{-}=\sigma_{+}^{\dagger}$ and $\sigma_{z}=2\sigma_{n}-\mathbb{I}$,
where $\sigma_{n}\equiv\sigma_{+}\sigma_{-}$. The relations $\sigma_{+}\sigma_{z}=\sigma_{+}=-\sigma_{z}\sigma_{+}$
and $\sigma_{z}\sigma_{-}=\sigma_{-}=-\sigma_{-}\sigma_{z}$ can also
be used. In this scheme, each spin orbital $i\uparrow$ is followed
in tensored space by the spin orbital $i\downarrow$. This ordering
is convenient to simplify the interaction terms of Fermi-Hubbard clusters
as the Coulomb interaction is confined to each site. As there is freedom
in the ordering of the indices, for other models a good ordering should
be chosen based on the symmetries of the simulated Hamiltonians. Note
that for finite clusters the Jordan-Wigner transformation is in general
independent of the dimensionality of the system.

\subsection{Measuring the correlation function\label{sub:MeasuringCorrelationFunction}}

\begin{figure}
\begin{centering}
\includegraphics[width=3.375in]{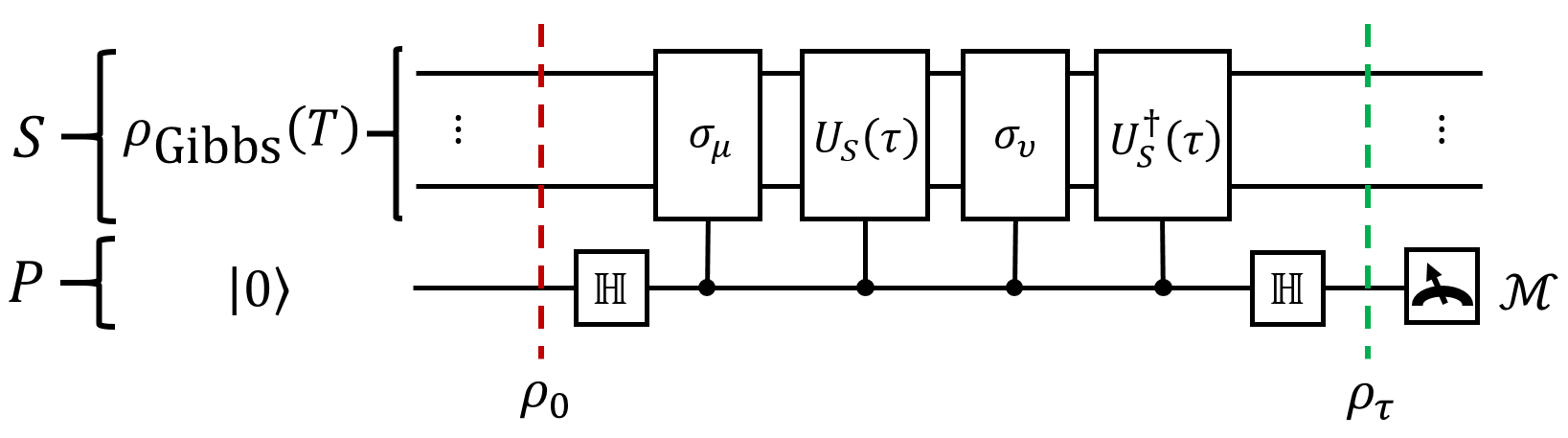}
\par\end{centering}

\caption{Circuit to measure the correlation function $C_{\mu\nu}\left(\tau\right)$
from an input Gibbs state as explained in Ref. \citep{DDW16}. Register
$S$ initially contains a given Gibbs state at inverse temperature
$\beta$ and register $P$ is a single qubit initialized in the zero
state. $P$ is put in a state superposition by applying a Hadamard
gate $\mathbb{H}$ and then used to apply the controlled evolution
sequence $O_{\mu\nu}\left(\tau\right)\equiv U_{S}^{\dagger}\left(\tau\right)\sigma_{\nu}U_{S}\left(\tau\right)\sigma_{\mu}$
with $U_{S}\left(\tau\right)=e^{-i\mathcal{H}'\tau}$ to the system
channel. Finally the state superposition is reversed by a last Hadamard
gate and the measurement in repeated to obtain the probability $P\left(\mathcal{M}\right)$,
which returns information on the cluster Green's function $\mathbf{\hat{G}'}\left(\omega\right)\equiv\left\langle \mathbf{\Psi}\mathbf{\Psi}^{\dagger}\right\rangle _{\omega}$.
\label{fig:CorrelationCircuit}}
\end{figure}
In figure \ref{fig:FullCircuit}, the Gibbs state produced in register
$S$ is conditionally evolved with gate $O\left(\tau\right)$ for
different times $\tau$ to measure the correlation functions. The
precise decomposition of $\mathrm{c-}O\left(\tau\right)$ in fermionic
operators and unitary Hamiltonian evolutions is shown in figure \ref{fig:CorrelationCircuit}
\citep{DDW16}. Registers $P$ and $S$ are initially in the separable
state $\left|0\right\rangle \left\langle 0\right|\otimes\rho_{\mathrm{Gibbs}}\left(T\right)$,
where
\begin{equation}
\rho_{\mathrm{Gibbs}}\left(T\right)\equiv\frac{1}{Z}\sum_{m}e^{-\frac{E_{m}}{T}}\left|\phi_{m}\right\rangle \left\langle \phi_{m}\right|\label{eq:GibbsState}
\end{equation}
and $E_{m}$ and $\left|\phi_{m}\right\rangle $ are respectively
the eigenenergies and eigenstates of $\mathcal{H}'$. A Hadamard gate
is applied on $P$ such that is is in the state $\frac{\left|0\right\rangle +\left|1\right\rangle }{\sqrt{2}}$.
Then a conditional Hermitianized creation/annihilation operator $\sigma_{\mu}\in\left\{ X_{i\sigma},Y_{i\sigma}\right\} $
is applied on register $S$ controlled from $P$. Since creation and
annihilation operators are not invertible, they cannot be used as
$\sigma_{\mu}$ and $\sigma_{\nu}$ directly. A trick consists in
using a linear combination of the operators. For each electron site,
the Hermitian $X_{i\sigma}$ and $Y_{i\sigma}$ operators are defined
from (\ref{eq:JordanWignerMap}) such that

\begin{equation}
\begin{array}{ccl}
X_{i\sigma} & \equiv & c_{i\sigma}+c_{i\sigma}^{\dagger}\\
\\
Y_{i\sigma} & \equiv & +i\left(c_{i\sigma}-c_{i\sigma}^{\dagger}\right).
\end{array}\label{eq:HermitianizedFermions}
\end{equation}
Note that $\left[X_{i\sigma},Y_{j\sigma'}\right]=i\delta_{ij}\delta_{\sigma\sigma'}Z_{i\sigma}$,
where $Z_{i\sigma}\equiv c_{i\sigma}^{\dagger}c_{i\sigma}-\frac{1}{2}$.
For a cluster with $L_{c}$ site, it is convenient to order the Jordan-Wigner
basis such that up/down spins orbitals for each site are adjacents:

\begin{equation}
\begin{array}{rcl}
X_{i\uparrow} & = & \mathbb{I}^{\otimes2\left(L_{c}-i\right)+1}\otimes\sigma_{x}\otimes\sigma_{z}^{\otimes2\left(i-1\right)}\\
\\
X_{i\downarrow} & = & \mathbb{I}^{\otimes2\left(L_{c}-i\right)}\otimes\sigma_{x}\otimes\sigma_{z}^{\otimes2i-1}\\
\\
Y_{i\uparrow} & = & \mathbb{I}^{\otimes2\left(L_{c}-i\right)+1}\otimes\sigma_{y}\otimes\sigma_{z}^{\otimes2\left(i-1\right)}\\
\\
Y_{i\downarrow} & = & \mathbb{I}^{\otimes2\left(L_{c}-i\right)}\otimes\sigma_{y}\otimes\sigma_{z}^{\otimes2i-1}.
\end{array}\label{eq:HermitianizedFermions1D}
\end{equation}
These operators must be implemented as operations controlled from
register $P$,
\begin{equation}
\begin{array}{rcl}
\mathrm{c-}X_{i\sigma} & = & \left|0\right\rangle \left\langle 0\right|\otimes\mathbb{I}^{\otimes2L_{c}}+\left|1\right\rangle \left\langle 1\right|\otimes X_{i\sigma}\\
\\
\mathrm{c-}Y_{i\sigma} & = & \left|0\right\rangle \left\langle 0\right|\otimes\mathbb{I}^{\otimes2L_{c}}+\left|1\right\rangle \left\langle 1\right|\otimes Y_{i\sigma}
\end{array}\label{eq:controlledHermitianizedFermions}
\end{equation}
for spins $\uparrow/\downarrow$ and $i$ between 1 and $L_{c}$.
These operators are easy to construct using the method found in \citep{Kaicher16}
and the types of sequences found in the next section.

Following the first $\mathrm{c-}\sigma_{\mu}$ operation, the $S$
register is conditionally evolved with the cluster Hamiltonian $\mathcal{H}'$:

\begin{equation}
\mathrm{c-}U_{S}\left(\tau\right)\equiv\left|0\right\rangle \left\langle 0\right|\otimes\mathbb{I}^{\otimes2L_{c}}+\left|1\right\rangle \left\langle 1\right|\otimes e^{-i\mathcal{H}'\tau}.\label{eq:controlledEvolution}
\end{equation}
Section \ref{sec:TimeEvolution} is dedicated to the precise gate
decompostion of (\ref{eq:controlledEvolution}) as it was treated
as a black-box in \citep{DDW16}.

After the second $\mathrm{c-}\sigma_{\nu}$ operation and the reverse
conditional time evolution is applied, register $P$ is measured and
register $S$ can be discarded. The measured probability outcomes
are recorded for each time $\tau$
\begin{equation}
C_{\mu\nu}\left(\tau\right)=2\left(P_{\mu\nu}\left(\mathcal{M}=0,\tau\right)-P_{\mu\nu}\left(\mathcal{M}=1,\tau\right)\right)\label{eq:MeasuredCorrelationFunction}
\end{equation}
such that the elements of the Nambu Green's function can be computed
from the inverse transformation

\begin{equation}
\resizebox{.9\hsize}{!}{\ensuremath{\left(\begin{array}{c}
\left\langle c_{i\sigma}\left(\tau\right)c_{j\sigma'}^{\dagger}\left(0\right)\right\rangle \\
\left\langle c_{i\sigma}^{\dagger}\left(\tau\right)c_{j\sigma'}\left(0\right)\right\rangle \\
\left\langle c_{i\sigma}^{\phantom{}}\left(\tau\right)c_{j\sigma'}\left(0\right)\right\rangle \\
\left\langle c_{i\sigma}^{\dagger}\left(\tau\right)c_{j\sigma'}^{\dagger}\left(0\right)\right\rangle 
\end{array}\right)=\frac{1}{2}\left(\begin{array}{rrrr}
1 & 1 & i & -i\\
1 & 1 & -i & i\\
1 & -1 & i & i\\
1 & -1 & -i & -i
\end{array}\right)\left(\begin{array}{c}
\left\langle X_{i\sigma}\left(\tau\right)X_{j\sigma'}\left(0\right)\right\rangle \\
\left\langle Y_{i\sigma}\left(\tau\right)Y_{j\sigma'}\left(0\right)\right\rangle \\
\left\langle Y_{i\sigma}\left(\tau\right)X_{j\sigma'}\left(0\right)\right\rangle \\
\left\langle X_{i\sigma}\left(\tau\right)Y_{j\sigma'}\left(0\right)\right\rangle 
\end{array}\right)}.}\label{eq:InverseJordanWigner}
\end{equation}
A simple Fourier transform then yield the retarded Green's function
$G_{\mu\nu}^{R}\left(\omega\right)$ which is used to iterated the
classical algorithm until a saddle-point $\frac{\partial\Omega_{t}}{\partial\mathbf{t'}}=0$
of the Potthoff self-energy function is found. Depending on the symmetries
of the cluster Hamiltonian, some terms in (\ref{eq:InverseJordanWigner})
may be zero at all time (e.g. if there is no pairing or spin-orbit
interaction) and can be removed from the computation for speed-up
or used to monitor possible errors coming from noise or other sources. 

Let's remark that (\ref{eq:MeasuredCorrelationFunction}) can be expanded
into a Taylor series

\begin{equation}
C_{\mu\nu}\left(\tau\right)=\sum_{s=0}^{\infty}\frac{\tau^{s}}{s!}C_{\mu\nu}^{\left(s\right)}.\label{eq:CorrelationFunctionMomentExpansion}
\end{equation}
The coefficients are also the moment of the Green's function in the
Lehmann representation such that the coefficient which can be measured
as the time derivative of (\ref{eq:MeasuredCorrelationFunction})
at $\tau\rightarrow0$: 

\begin{equation}
\begin{array}{rcl}
C_{\mu\nu}^{\left(s\right)} & = & \left(-i\right)^{s}\sum_{m}\sum_{n}A_{\mu\nu}^{mn}\left(E_{m}-E_{n}\right)^{s}\\
\\
 & = & \underset{\tau\rightarrow0^{+}}{\mathrm{lim}}\frac{d^{s}}{d\tau^{s}}C_{\mu\nu}\left(\tau\right)
\end{array}\label{eq:CorrelationFunctionMoments}
\end{equation}
The retarded Green's function is then given by 
\begin{equation}
G_{\mu\nu}^{R}\left(\omega\right)=-i\lim_{\eta\rightarrow0^{+}}\sum_{s=0}^{\infty}\frac{C_{\mu\nu}^{\left(s\right)}}{\left(\eta+i\omega\right)^{s+1}},\label{eq:RetardedGreensFunctionMoments}
\end{equation}
where $\eta$ is the small parameter of the analytical continuation
of the retarded function. In practice it can also be seen as an effective
inverse simulated time (or ``decoherence rate''). If one can measure
several cycles of the correlation functions (\ref{eq:CorrelationFunctionMomentExpansion}),
then the extracted spectra will be sharply defined and $\eta$ can
be considered effectively small with respect to all simulated energies
in the cluster Hamiltonian. In the other limit, if there is too much
decoherence in the quantum simulator the measured correlation functions
will be flat and no information can be extracted about the frequency
dependence of (\ref{eq:RetardedGreensFunctionMoments}), $\eta$ is
then effectively related to the decoherence rate if it limits the
simulated time.

\section{Time evolution of the cluster Hamiltonian\label{sec:TimeEvolution}}

In this section we will show how a typical trial cluster Hamiltonian
for the Fermi-Hubbard model in 2D can be implemented accurately using
a reasonable number of gates. In order to keep the notation straightforward,
this is done through the example of a $2\times2$ cluster with magnetic
and superconducting trial terms which can be easily generalized to
larger sizes and higher dimensions. After introducing the cluster
Hamiltonian and some notation, the gates for the implementation of
(\ref{eq:controlledEvolution}) will be shown for the example and
a numerical estimate of the Trotter-Suzuki error will be provided.
Along the way, ``conditional imaginary swaps'' or $\mathrm{c-\pm iSWAP}$s
will be introduced as three-qubit quantum gates pratical for quantum
simulations. Although they can be viewed as a complements to the traditional
$\mathrm{Toffoli}$ ($\mathrm{c-c-NOT}$) and $\mathrm{Fredkin}$
($\mathrm{c-SWAP}$) gates \citep{Nielsen01}, the positive or negative
imaginary phase in the ``$\mathrm{\pm iFredkin}$'' gates has no
classical analog and makes them truly quantum operations.

\subsection{Hamiltonian of a cluster \label{sub:HamiltonianCluster}}

Each cluster includes only a small subset of the terms of the original
lattice and variational terms must also be included to account for
possible long-range order. For convenience, let's assume a square
lattice with constant spacing $a$. It is broken down into $N_{c}$
clusters each with $L_{c}$ $s$-shell sites with two electrons each
(spin up $\uparrow$ and spin down $\downarrow$). The Hamiltonian
of each cluster is given by

\begin{equation}
\mathcal{H}'=\mathcal{H}_{\mathrm{kin}}+\mathcal{H}_{\mathrm{int}}-\mathcal{H}_{\mathrm{s-pair}}-\mathcal{H}_{\mathrm{d_{x^{2}-y^{2}}}}-\mathcal{H}_{\mathrm{local}}-\mathcal{H}_{\mathrm{AF}},\label{eq:HamiltonianOneCluster}
\end{equation}
where $\mathcal{H}_{\mathrm{kin}}$ is the kinetic term, $\mathcal{H}_{\mathrm{int}}$
is the local Coulomb interaction, $\mathcal{H}_{\mathrm{s-pair}}$
and $\mathcal{H}_{\mathrm{d_{x^{2}-y^{2}}}}$ are variational pairing
terms, $\mathcal{H}_{\mathrm{local}}$ is a variational chemical potential
term and $\mathcal{H}_{\mathrm{AF}}$ is a variational Néel antiferromagnetic
term. The variational self-energy functional method support many different
Hamiltonian terms and models as long as the two-body interaction term
is ``local'' enough that a cluster decomposition can be made without
cutting any interaction link.

\begin{figure}
\begin{centering}
\includegraphics[width=3.375in]{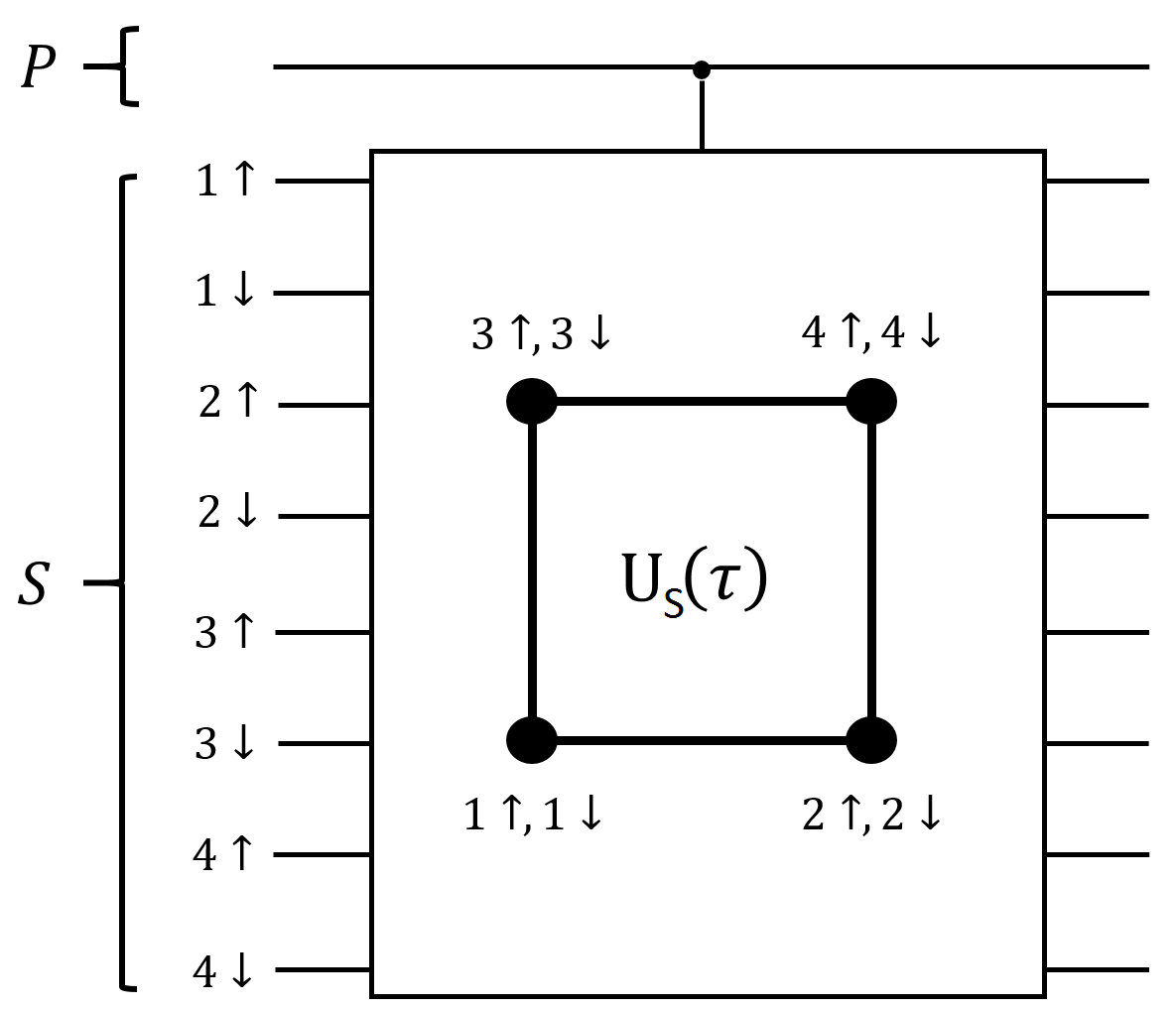}
\par\end{centering}

\caption{The chain of physical qubits representing the system register is easy
to represent and operate on in the gate model of computation. For
a $n=2L_{c}$ square lattice, the sites are labelled sequentially
in linear stripes, this ensures that nearest-neighor coupling terms
of the Hamiltonian in the Jordan-Wigner basis can be represented as
Pauli string of length at most $O\left(2\sqrt{L_{c}}\right)$.\label{fig:2x2Mapping}}
\end{figure}

Figure \ref{fig:2x2Mapping} show how the qubits of $S$ register
are labelled to represent the electronic structure of the cluster
and requires $2L_{c}$ qubits (1 qubit = 1 spin-orbital). Since the
qubits are effectively distinguishable spins, the Jordan-Wigner transformation
from section \ref{sub:JordanWigneTransformation} must be used to
model accurately the fermionic statistics of indistinguishable electrons.
The sites are simply assumed to be labelled sequentially when counting
the gate numbers for larger cluster sizes in section \ref{sec:Scaling}.

\begin{widetext}

\subsubsection{Some convenient Pauli strings}

To define the Hamiltonian terms of (\ref{eq:HamiltonianOneCluster})
in the Jordan-Wigner basis, it is useful to introduce the following
strings of Pauli matrices. The hopping part of the Hamiltonian usually
contains terms of the form

\begin{equation}
\begin{array}{rcl}
\mathbb{T}_{L_{c}\uparrow}\left(i,j\right) & \equiv & \mathbb{I}^{\otimes2\left(L_{c}-j\right)+1}\otimes\left(\sigma_{+}\otimes\sigma_{z}^{\otimes2\left(j-i\right)-1}\otimes\sigma_{-}+\sigma_{-}\otimes\sigma_{z}^{\otimes2\left(j-i\right)-1}\otimes\sigma_{+}\right)\otimes\mathbb{I}^{\otimes2\left(i-1\right)}\\
 & = & 2\mathbb{I}^{\otimes2\left(L_{c}-j\right)+1}\otimes\left(\sigma_{x}\otimes\sigma_{z}^{\otimes2\left(i-j\right)-1}\otimes\sigma_{x}+\sigma_{y}\otimes\sigma_{z}^{\otimes2\left(j-i\right)-1}\otimes\sigma_{y}\right)\otimes\mathbb{I}^{\otimes2\left(i-1\right)}\\
\\
\mathbb{T}_{L_{c}\downarrow}\left(i,j\right) & \equiv & \mathbb{I}^{\otimes2\left(L_{c}-j\right)}\otimes\left(\sigma_{+}\otimes\sigma_{z}^{\otimes2\left(i-j\right)-1}\otimes\sigma_{-}+\sigma_{-}\otimes\sigma_{z}^{\otimes2\left(j-i\right)-1}\otimes\sigma_{+}\right)\otimes\mathbb{I}^{\otimes2i-1}\\
 & = & 2\mathbb{I}^{\otimes2\left(L_{c}-j\right)}\otimes\left(\sigma_{x}\otimes\sigma_{z}^{\otimes2\left(i-j\right)-1}\otimes\sigma_{x}+\sigma_{y}\otimes\sigma_{z}^{\otimes2\left(j-i\right)-1}\otimes\sigma_{y}\right)\otimes\mathbb{I}^{\otimes2i-1},
\end{array}\label{eq:KinOperator}
\end{equation}
where $j>i$ between 1 and $L_{c}$. These strings have the property
$\left[\mathbb{T}_{L_{c}\sigma}\left(i,j\right),\mathbb{T}_{L_{c}\sigma'}\left(i',j'\right)\right]=0$.
The chemical potential and the variational antiferromagnetic terms
built from $n_{i\sigma}$ operators have strings of the form

\begin{equation}
\begin{array}{rcl}
\mathbb{T}_{L_{c}\uparrow}\left(i\right) & \equiv & \mathbb{I}^{\otimes2\left(L_{c}-i\right)+1}\otimes\sigma_{n}\otimes\mathbb{I}^{\otimes2\left(i-1\right)}\\
\\
\mathbb{T}_{L_{c}\downarrow}\left(i\right) & \equiv & \mathbb{I}^{\otimes2\left(L_{c}-i\right)}\otimes\sigma_{n}\otimes\mathbb{I}^{\otimes2i-1}.
\end{array}\label{eq:LocOperator}
\end{equation}
Since $\mathbb{T}_{L_{c}}\left(i,j\right)$ and $\mathbb{T}_{L_{c}}\left(i\right)$
conserve total spin in the Pauli basis, they are also number conserving
in the occupation basis.

\begin{equation}
\begin{array}{rcl}
\mathbb{D}_{L_{c}\uparrow}\left(i,j\right) & \equiv & \mathbb{I}^{\otimes2\left(L_{c}-j\right)}\otimes\left(\sigma_{+}\otimes\sigma_{z}^{\otimes2\left(j-i\right)}\otimes\sigma_{+}+\sigma_{-}\otimes\sigma_{z}^{\otimes2\left(j-i\right)}\otimes\sigma_{-}\right)\otimes\mathbb{I}^{\otimes2\left(i-1\right)}\\
 & = & 2\mathbb{I}^{\otimes2\left(L_{c}-j\right)}\otimes\left(\sigma_{x}\otimes\sigma_{z}^{\otimes2\left(j-i\right)}\otimes\sigma_{x}-\sigma_{y}\otimes\sigma_{z}^{\otimes2\left(j-i\right)}\otimes\sigma_{y}\right)\otimes\mathbb{I}^{\otimes2\left(i-1\right)}\\
\\
\mathbb{D}_{L_{c}\downarrow}\left(i,j\right) & \equiv & \mathbb{I}^{\otimes2\left(L_{c}-j\right)+1}\otimes\left(\sigma_{+}\otimes\sigma_{z}^{\otimes2\left(j-i-1\right)}\otimes\sigma_{+}+\sigma_{-}\otimes\sigma_{z}^{\otimes2\left(j-i-1\right)}\otimes\sigma_{-}\right)\otimes\mathbb{I}^{\otimes2i-1}\\
 & = & 2\mathbb{I}^{\otimes2\left(L_{c}-j\right)+1}\otimes\left(\sigma_{x}\otimes\sigma_{z}^{\otimes2\left(j-i-1\right)}\otimes\sigma_{x}-\sigma_{y}\otimes\sigma_{z}^{\otimes2\left(j-i-1\right)}\otimes\sigma_{y}\right)\otimes\mathbb{I}^{\otimes2i-1}
\end{array}\label{eq:PairOperator}
\end{equation}
in this case, $j>i$ can be anything between 1 and $L_{c}$. $\mathbb{D}_{L_{c}}\left(i,j\right)$
does not conserve total spin in the Pauli basis and it is not conserving
in the occupation basis. The $\mathbb{D}_{L_{c}\sigma}\left(i,j\right)$
are used to represent pairing operators between differents sites in
the Pauli basis.

\begin{equation}
\begin{array}{rcl}
\mathbb{D}_{L_{c}}\left(i\right) & \equiv & \mathbb{I}^{\otimes2\left(L_{c}-i\right)}\otimes\left(\sigma_{+}\otimes\sigma_{+}+\sigma_{-}\otimes\sigma_{-}\right)\otimes\mathbb{I}^{\otimes2\left(i-1\right)}\\
 & = & 2\mathbb{I}^{\otimes2\left(L_{c}-i\right)}\otimes\left(\sigma_{x}\otimes\sigma_{x}-\sigma_{y}\otimes\sigma_{y}\right)\otimes\mathbb{I}^{\otimes2\left(i-1\right)}
\end{array}\label{eq:LocPairOperator}
\end{equation}
The $\mathbb{D}_{L_{c}}\left(i\right)$ operators are used to represent
local pairing operators in the Pauli basis.\end{widetext}

\subsection{Gate decomposition\label{sub:GateDecomposition}}

Here we proceed to decomposing the terms of the cluster Hamiltonian
(\ref{eq:HamiltonianOneCluster}). This is not an exhaustive list
of all possible variational terms nor of the detailed decomposition
method as it is covered in \citep{Kaicher16}. The aim is to provide
an estimate of the number of quantum gates required during the simulation
of the Fermi-Hubbard model. It is also shown that different blocks
of the cluster Hamiltonian can be implemented exactly. The time evolution
of the blocks that do not commute can be approximated by a Trotter-Suzuki
approximation detailed in section (\ref{sub:TrotterSuzuki}).

Let's note we are using $\mathbb{H}=\frac{1}{\sqrt{2}}\left(\begin{array}{cc}
1 & 1\\
1 & -1
\end{array}\right)$ and $\mathbb{J}=\frac{1}{\sqrt{2}}\left(\begin{array}{cc}
1 & -i\\
1 & i
\end{array}\right)$. Given a tunable nearest-neighbor exchange interaction $\sigma_{x}\otimes\sigma_{x}+\sigma_{y}\otimes\sigma_{y}$
between the qubits of register $S$, it naturally generates the ``imaginary
swap'' gate

\begin{equation}
\begin{array}{rcl}
\mathrm{\pm iSWAP} & = & e^{\pm i\frac{\pi}{4}\left(\sigma_{x}\otimes\sigma_{x}+\sigma_{y}\otimes\sigma_{y}\right)}\\
\\
 & = & \left(\begin{array}{cccc}
1 & 0 & 0 & 0\\
0 & 0 & \pm i & 0\\
0 & \pm i & 0 & 0\\
0 & 0 & 0 & 1
\end{array}\right).
\end{array}\label{eq:iSWAP}
\end{equation}
It has the nice property that it can be used to manipulate Pauli strings
that appear in the Jordan-Wigner representation:

\begin{equation}
\begin{array}{ccc}
\mathrm{+iSWAP}\cdot\left(\mathbb{I}\otimes\sigma_{x}\right)\cdot\mathrm{-iSWAP} & = & \sigma_{y}\otimes\sigma_{z}\\
\\
\mathrm{+iSWAP}\cdot\left(\mathbb{I}\otimes\sigma_{y}\right)\cdot\mathrm{-iSWAP} & = & -\sigma_{x}\otimes\sigma_{z}\\
\\
\mathrm{+iSWAP}\cdot\left(\mathbb{I}\otimes\sigma_{z}\right)\cdot\mathrm{-iSWAP} & = & \sigma_{z}\otimes\mathbb{I}.
\end{array}\label{eq:PauliStringsiSWAP}
\end{equation}
To implement a conditional evolution gates of the form (\ref{eq:controlledEvolution}),
we introduce $\mathrm{c-\pm iSWAP}$s as fundamental 3-qubit gates
for quantum simulations. These gates come only in two varieties ($\pm$)
for each triple of qubits (qubit $P$ and two adjacent qubits in $S$).
Since all other operations are conditional single-qubit gates, they
are expected to be the most time-consuming operations and therefore
they are used to benchmark the scaling properties of the algorithm.
Let's note that there appears to be numerical evidence that coupling
the $P$ and the $S$ registers with tunable $\sigma_{z}\otimes\sigma_{z}$
interactions greatly simplifies the implementation of the $\mathrm{c-\pm iSWAP}$
gates\citep{Liebermann16}. This somewhat extends the toolset of three-qubit
gates for reversible quantum computation, which already contains Toffoli
and Fredkin gates. ``Conditional single-qubit gates'' is abbreviated
by $\mathrm{c-SQG}$.

\subsubsection{Local terms}

\begin{figure*}
\begin{centering}
\begin{minipage}[t]{1\columnwidth}%
\begin{center}
\Qcircuit @C=0.2em @R=.2em @!R { \lstick{P}					& \ctrl{1} & \ctrl{2} & \ctrl{3} & \ctrl{4} & \ctrl{5} & \ctrl{6} & \ctrl{7} & \ctrl{8} & \qw\\ \lstick{1\uparrow}	& \gate{R_{\sigma_n}^{\Theta_{\rm loc}^+}} & \qw & \qw & \qw & \qw & \qw & \qw & \qw & \qw \\ \lstick{1\downarrow}& \qw & \gate{R_{\sigma_n}^{\Theta_{\rm loc}^-}} & \qw & \qw & \qw & \qw & \qw & \qw & \qw \\ \lstick{2\uparrow}	& \qw & \qw & \gate{R_{\sigma_n}^{\Theta_{\rm loc}^-}} & \qw & \qw & \qw & \qw & \qw & \qw \\ \lstick{2\downarrow}& \qw & \qw & \qw & \gate{R_{\sigma_n}^{\Theta_{\rm loc}^+}} & \qw & \qw & \qw & \qw & \qw \\ \lstick{3\uparrow}	& \qw & \qw & \qw & \qw & \gate{R_{\sigma_n}^{\Theta_{\rm loc}^-}} & \qw & \qw & \qw & \qw \\ \lstick{3\downarrow}& \qw & \qw & \qw & \qw & \qw & \gate{R_{\sigma_n}^{\Theta_{\rm loc}^+}} & \qw & \qw & \qw \\ \lstick{4\uparrow}	& \qw & \qw & \qw & \qw & \qw & \qw & \gate{R_{\sigma_n}^{\Theta_{\rm loc}^+}} & \qw & \qw \\ \lstick{4\downarrow}& \qw & \qw & \qw & \qw & \qw & \qw & \qw & \gate{R_{\sigma_n}^{\Theta_{\rm loc}^-}} & \qw }
\par\end{center}%
\end{minipage}
\par\end{centering}

\centering{}\caption{The local terms of the cluster Hamiltonian corresponding to the time
evolution of $\mathcal{H}_{\mathrm{local}}$ and $\mathcal{H}_{\mathrm{AF}}$.
The single qubit rotation $R_{\sigma_{n}}^{\Theta}\equiv e^{-i\frac{\Theta}{2}}e^{-i\frac{\Theta}{2}\sigma_{z}}$,
the angles $\Theta_{{\rm loc}}^{\pm}\equiv-\Delta\tau\left(\mu'\pm M'\right)$.
There are $2L_{c}$ $\mathrm{c-SQG}$s in a square cluster (8 $\mathrm{c-SQG}$s
in a $2\times2$ cluster).\label{fig:SimulationLocalTerms}}
\end{figure*}
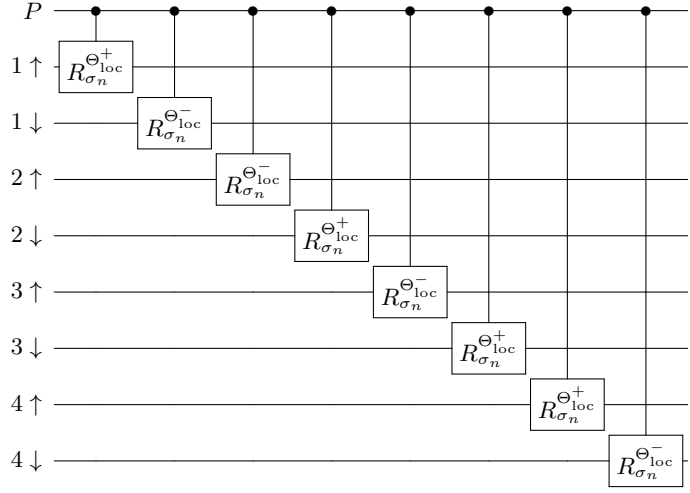
Local terms are all one-body terms composed with the $n_{i\sigma}$
operators. This includes the chemical potential

\begin{equation}
\begin{array}{rcl}
\mathcal{H}_{\mathrm{local}} & = & \mu'\sum_{i,\sigma}n_{i\sigma}\\
\\
 & = & \mu'\sum_{i=1}^{L_{c}}\left(\mathbb{T}_{L_{c}\uparrow}\left(i\right)+\mathbb{T}_{L_{c}\downarrow}\left(i\right)\right)
\end{array}\label{eq:VariationChemicalPotential}
\end{equation}
which is kept as a variational term to enforce the thermodynamic consistency
of the electronic occupation value. The $\mathbb{T}_{L_{c}\sigma}\left(i\right)$
strings are given by (\ref{eq:LocOperator}). The variational Néel
antiferromagnetic Weiss field is also a local term which takes the
form 
\begin{equation}
\begin{array}{rcl}
\mathcal{H}_{\mathrm{AF}} & = & M'\sum_{i}e^{i\mathbf{Q}\cdot\mathbf{R}_{i}}\left(n_{i\uparrow}-n_{i\downarrow}\right)\\
\\
 & = & M'\sum_{i=1}^{L_{c}}e^{i\mathbf{Q}\cdot\mathbf{R}_{i}}\left(\mathbb{T}_{L_{c}\uparrow}\left(i\right)-\mathbb{T}_{L_{c}\downarrow}\left(i\right)\right)
\end{array}\label{eq:VariationalAF}
\end{equation}
where $\mathbf{Q}=\left(\pi,\pi\right)$ is the antiferromagnetic
wavevector and $\mathbf{R}_{i}$ is the position of the site in units
of $a$. These terms all commute between each other and do not require
any $\mathrm{c-\pm iSWAP}$, only 2$L_{c}$ $\mathrm{c-}R_{\sigma_{n}}^{\Theta}$
are required, where
\begin{equation}
R_{\sigma_{n}}^{\Theta}\equiv e^{-i\frac{\Theta}{2}}e^{-i\frac{\Theta}{2}\sigma_{z}}.\label{eq:NumberRotation}
\end{equation}
The gate sequence is shown in figure \ref{fig:SimulationLocalTerms}.

\subsubsection{Interaction terms}

\begin{figure*}
\begin{centering}
\begin{minipage}[t]{1\columnwidth}%
\begin{center}
\subfigure[]{\label{fig:SimulationInteractionTerms}\Qcircuit @C=1em @R=.7em @!R { \lstick{P}					& \ctrl{1} & \ctrl{3} & \ctrl{5} & \ctrl{7} & \qw\\ \lstick{1\uparrow}	& \multigate{1}{R_{U}^{\Theta_{\rm int}}} & \qw & \qw & \qw & \qw \\ \lstick{1\downarrow}& \ghost{R_{\sigma_n}^{\Theta_{\rm loc}^-}} & \qw & \qw & \qw & \qw \\ \lstick{2\uparrow}	& \qw & \multigate{1}{R_{U}^{\Theta_{\rm int}}} & \qw & \qw & \qw \\ \lstick{2\downarrow}& \qw & \ghost{R_{\sigma_n}^{\Theta_{\rm loc}^+}} & \qw & \qw & \qw \\ \lstick{3\uparrow}	& \qw & \qw & \multigate{1}{R_{U}^{\Theta_{\rm int}}} & \qw & \qw \\ \lstick{3\downarrow}& \qw & \qw & \ghost{R_{\sigma_n}^{\Theta_{\rm loc}^+}} & \qw & \qw \\ \lstick{4\uparrow}	& \qw & \qw & \qw & \multigate{1}{R_{U}^{\Theta_{\rm int}}} & \qw \\ \lstick{4\downarrow}& \qw & \qw & \qw & \ghost{R_{\sigma_n}^{\Theta_{\rm loc}^-}} & \qw \\ & & & & &}}
\par\end{center}%
\end{minipage}\vfill{}
\begin{minipage}[t]{1\columnwidth}%
\begin{center}
\subfigure[]{\label{fig:InteractionRotationGate}\Qcircuit @C=1em @R=.7em @!R { \lstick{P} 						& \ctrl{1} 										& \qw & 																					& & \lstick{P}				 		& \ctrl{2} & \ctrl{1} & \ctrl{1} & \ctrl{1} & \ctrl{2} & \ctrl{2} & \ctrl{1} & \qw\\ \lstick{i\uparrow} 		& \multigate{1}{R_U^{\Theta}}	& \qw & \push{\rule{.3em}{0em}=\rule{.3em}{0em}}	& & \lstick{i\uparrow}		& \qw & \multigate{1}{-} & \gate{R_{\sigma_n}^{\frac{\Theta}{2}}} & \multigate{1}{+} & \qw & \qw & \gate{R_{\sigma_n}^{\frac{\Theta}{2}}} & \qw\\ \lstick{i\downarrow}	& \ghost{R_U^{\Theta}} 				& \qw & 																					& & \lstick{i\downarrow}	& \gate{\mathbb{H}} & \ghost{-} & \qw & \ghost{+} & \gate{\mathbb{H}} & \gate{R_{\sigma_n}^{\frac{\Theta}{2}}} & \qw & \qw \\ & & & & & & & & & & & & &}}
\par\end{center}%
\end{minipage}
\par\end{centering}

\centering{}\caption{In \subref{fig:SimulationInteractionTerms}, the interaction terms
of the cluster Hamiltonian corresponding to the time evolution of
$\mathcal{H_{\mathrm{int}}}$ are decomposed into gates. The angle
$\Theta_{{\rm int}}\equiv+\Delta\tau U$. In \subref{fig:InteractionRotationGate},
the decomposition of $\mathrm{c-}R_{U}^{\Theta}$ in a site subspace
(spin $\uparrow$/$\downarrow$) is shown. There are $L_{c}$ terms
like these in a square cluster. The single-qubit rotation gate $R_{\sigma_{U}}^{\Theta}\equiv e^{+i\frac{\Theta}{2}}e^{-i\frac{\Theta}{2}\sigma_{y}}$.
There are 5 $\mathrm{c-SQG}$s and 2 $\mathrm{c-\pm iSWAP}$s per
$\mathrm{c-}R_{U}^{\Theta}$.\label{fig:Interaction}}
\end{figure*}
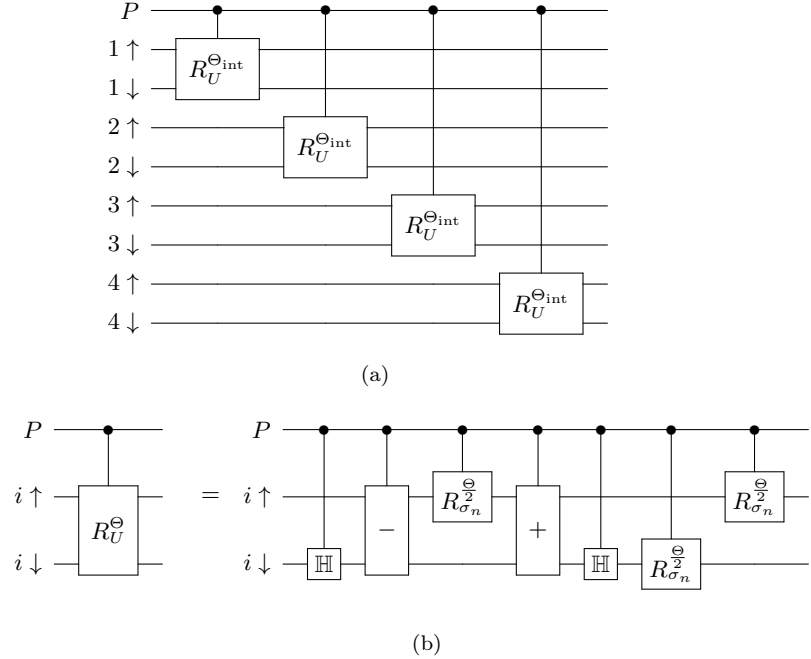
The fixed interaction terms are given by
\begin{equation}
\begin{array}{rcl}
\mathcal{H_{\mathrm{int}}} & = & U\sum_{i}n_{i\uparrow}n_{i\downarrow}\\
\\
 & = & U\sum_{i=1}^{L_{c}}\mathbb{T}_{L_{c}\uparrow}\left(i\right)\cdot\mathbb{T}_{L_{c}\downarrow}\left(i\right),
\end{array}\label{eq:InteractionHamiltonianCluster}
\end{equation}
where the $\mathbb{T}_{L_{c}\sigma}\left(i\right)$ strings are given
by (\ref{eq:LocOperator}). From figures \ref{fig:SimulationInteractionTerms}
and \ref{fig:InteractionRotationGate}, it can seen that $L_{c}$
$\mathrm{c-+iSWAP}$s, $L_{c}$ $\mathrm{c--iSWAP}$s , $2L_{c}$
$\mathrm{c-}\mathbb{H}$ on spin-$\downarrow$ orbitals, $L_{c}$
$\mathrm{c-}R_{\sigma_{U}}^{\Theta}$ on spin-$\uparrow$ orbitals
and $2L_{c}$$\mathrm{c-}R_{\sigma_{n}}^{\Theta}$ on all qubits (those
should be done at the same time as the gates of figure \ref{fig:SimulationLocalTerms},
then only the ressources from the interaction terms have to be counted)
are required to implement the evolution of $\mathcal{H_{\mathrm{int}}}+\mathcal{H_{\mathrm{local}}}+\mathcal{H_{\mathrm{AF}}}$.
These term are simple to implement and they commute with the local
terms $\mathcal{H}_{\mathrm{local}}$ and $\mathcal{H}_{\textrm{AF}}$,
so they should be done in sequence.

\subsubsection{Hopping terms}

\begin{figure*}
\begin{centering}
\begin{minipage}[t]{1\columnwidth}%
\begin{center}
\subfigure[]{\label{fig:SimulationHoppingTerms}\Qcircuit @C=1em @R=.7em @!R { \lstick{P}					& \ctrl{1}  & \ctrl{1} & \ctrl{3} & \ctrl{5} & \ctrl{2} & \ctrl{2} & \ctrl{4} & \ctrl{6} & \qw\\ \lstick{1\uparrow}	& \multigate{2}{R_{K1}^{\Theta_{\rm kin}}} & \multigate{4}{R_{K2}^{\Theta_{\rm kin}}} & \qw & \qw & \qw & \qw & \qw & \qw & \qw \\ \lstick{1\downarrow}& \ghost{R_{K1}^{\Theta_{\rm kin}}} & \ghost{R_{K2}^{\Theta_{\rm kin}}} & \qw & \qw & \multigate{2}{R_{K1}^{\Theta_{\rm kin}}} & \multigate{4}{R_{K2}^{\Theta_{\rm kin}}} & \qw & \qw & \qw \\ \lstick{2\uparrow}	& \ghost{R_{K1}^{\Theta_{\rm kin}}} & \ghost{R_{K2}^{\Theta_{\rm kin}}} & \multigate{4}{R_{K2}^{\Theta_{\rm kin}}} & \qw & \ghost{R_{K1}^{\Theta_{\rm kin}}} & \ghost{R_{K2}^{\Theta_{\rm kin}}} & \qw & \qw & \qw \\ \lstick{2\downarrow}& \qw & \ghost{R_{K2}^{\Theta_{\rm kin}}} & \ghost{R_{K2}^{\Theta_{\rm kin}}} & \qw & \ghost{R_{K1}^{\Theta_{\rm kin}}} & \ghost{R_{K2}^{\Theta_{\rm kin}}} & \multigate{4}{R_{K2}^{\Theta_{\rm kin}}} & \qw & \qw \\ \lstick{3\uparrow}	& \qw & \ghost{R_{K2}^{\Theta_{\rm kin}}} & \ghost{R_{K2}^{\Theta_{\rm kin}}} & \multigate{2}{R_{K1}^{\Theta_{\rm kin}}} & \qw & \ghost{R_{K2}^{\Theta_{\rm kin}}} & \ghost{R_{K2}^{\Theta_{\rm kin}}} & \qw & \qw \\ \lstick{3\downarrow}& \qw & \qw & \ghost{R_{K2}^{\Theta_{\rm kin}}} & \ghost{R_{K1}^{\Theta_{\rm kin}}} & \qw & \ghost{R_{K2}^{\Theta_{\rm kin}}} & \ghost{R_{K2}^{\Theta_{\rm kin}}} & \multigate{2}{R_{K1}^{\Theta_{\rm kin}}} & \qw \\ \lstick{4\uparrow}	& \qw & \qw & \ghost{R_{K2}^{\Theta_{\rm kin}}} & \ghost{R_{K1}^{\Theta_{\rm kin}}} & \qw & \qw & \ghost{R_{K2}^{\Theta_{\rm kin}}} & \ghost{R_{K1}^{\Theta_{\rm kin}}} & \qw \\ \lstick{4\downarrow}& \qw & \qw & \qw & \qw & \qw & \qw & \ghost{R_{K2}^{\Theta_{\rm kin}}} & \ghost{R_{K1}^{\Theta_{\rm kin}}} & \qw \\ & & & & & & & & &}}
\par\end{center}%
\end{minipage}\vfill{}
\begin{minipage}[t]{1\columnwidth}%
\begin{center}
\subfigure[]{\label{fig:HoppingRotationGate}\Qcircuit @C=0.2em @R=.5em @!R { \lstick{P}															& \ctrl{1} 												& \qw & 																					& & \ctrl{1} & \ctrl{5} & \push{\cdots} \qw & \ctrl{3} & \ctrl{2} & \ctrl{1} & \ctrl{1} & \ctrl{1} & \ctrl{1} & \ctrl{1} & \ctrl{1} & \ctrl{1} & \ctrl{1} & \ctrl{2} & \ctrl{3} & \push{\cdots} \qw & \ctrl{5} & \ctrl{1} & \qw \\ \lstick{i\uparrow/i\downarrow}					& \multigate{5}{R_{Km}^{\Theta}}	& \qw &																						& & \gate{\mathbb{H}} & \qw & \qw & \qw & \qw & \multigate{1}{-} & \gate{R_{\sigma_x}^{2\Theta}} & \multigate{1}{+} & \gate{\mathbb{H}} & \gate{\mathbb{J}} & \multigate{1}{-} & \gate{R_{\sigma_y}^{2\Theta}} & \multigate{1}{+} & \qw & \qw & \qw & \qw & \gate{\mathbb{J}^\dagger} & \qw \\ \lstick{i\downarrow/(i+1)\uparrow}			& \ghost{R_{Km}^{\Theta}}					& \qw &																						& & \qw & \qw & \qw & \qw & \multigate{1}{+} & \ghost{-} & \qw & \ghost{+} & \qw & \qw & \ghost{-} & \qw & \ghost{+} & \multigate{1}{-} & \qw & \qw & \qw & \qw & \qw \\ \lstick{(i+1)\uparrow/(i+1)\downarrow}	& \ghost{R_{Km}^{\Theta}}					& \qw & \push{\rule{.3em}{0em}=\rule{.3em}{0em}}	& & \qw & \qw & \qw & \multigate{1}{-} & \ghost{+} & \qw & \qw & \qw & \qw & \qw & \qw & \qw & \qw & \ghost{-} & \multigate{1}{+} & \qw & \qw & \qw & \qw \\ \lstick{\vdots}	 												& \ghost{R_{Km}^{\Theta}}					& \qw &																						& & \qw & \qw & \push{\iddots} \qw & \ghost{-} & \qw & \qw & \qw & \qw & \qw & \qw & \qw & \qw & \qw & \qw & \ghost{+} & \push{\ddots}\qw & \qw & \qw & \qw \\ \lstick{\vdots}													& \ghost{R_{Km}^{\Theta}}					& \qw &																						& & \qw & \multigate{1}{+} & \push{\iddots} \qw & \qw & \qw & \qw & \qw & \qw & \qw & \qw & \qw & \qw & \qw & \qw & \qw & \push{\ddots}\qw & \multigate{1}{-} & \qw & \qw \\ \lstick{(i+m)\uparrow/(i+m)\downarrow}	& \ghost{R_{Km}^{\Theta}}					& \qw &																						& & \qw & \ghost{+} & \qw & \qw & \qw & \qw & \qw & \qw & \qw & \qw & \qw & \qw & \qw & \qw & \qw & \qw & \ghost{-} & \qw & \qw \\ & & & & & & & & & & & & & & & & & & & & & & &}}
\par\end{center}%
\end{minipage}
\par\end{centering}

\caption{In \subref{fig:SimulationHoppingTerms}, the hopping terms of the
cluster Hamiltonian corresponding to the time evolution of $\mathcal{H_{\mathrm{kin}}}$
are decomposed into gates. The angle $\Theta_{{\rm int}}\equiv-\Delta\tau t$.
There are $4\left(L_{c}-\sqrt{L_{c}}\right)$ terms like these in
a square lattice. Half contains Pauli strings of length 3 and the
other half has length $2\sqrt{L_{c}}+1$. In \subref{fig:HoppingRotationGate},
the decomposition of $\mathrm{c-}R_{Km}^{\Theta}$ in a subspace starting
at $i\uparrow$ ($i\downarrow$) and ending at $i+m\uparrow$ ($i+m\downarrow$),
where $m=1$ or $\sqrt{L_{c}}$ in a square lattice with nearest-neighbor
hopping. There are 6 $\mathrm{c-SQG}$s and $4m$ $\mathrm{c-\pm iSWAP}$s
per $\mathrm{c-}R_{Km}^{\Theta}$.\label{fig:Hopping}}

\end{figure*}
The hopping terms between nearest-neighbors is given by 
\begin{equation}
\begin{array}{rcl}
\mathcal{H_{\mathrm{kin}}} & = & -t\sum_{\left\langle i,j\right\rangle ,\sigma}c_{i\sigma}^{\dagger}c_{j\sigma}+c_{j\sigma}^{\dagger}c_{i\sigma}\\
\\
 & = & -t\sum_{\left\langle i,j\right\rangle }\left(\mathbb{T}_{L_{c}\uparrow}\left(i,j\right)+\mathbb{T}_{L_{c}\downarrow}\left(i,j\right)\right)
\end{array}\label{eq:KinecticHamiltonianCluster}
\end{equation}
for all neighboring orbitals $\left\langle i,j\right\rangle $ such
that $j>i$. The summation $\sum_{\left\langle i,j\right\rangle }$
has $2\left(L_{c}-\sqrt{L_{c}}\right)$ nearest-neighbor vertices.
The $\mathbb{T}_{L_{c}\sigma}\left(i,j\right)$ strings are given
by (\ref{eq:KinOperator}). From figures \ref{fig:SimulationHoppingTerms}
and \ref{fig:HoppingRotationGate}, it can seen that $4\left(\sqrt[3]{L_{c}}-\sqrt{L_{c}}\right)$
$\mathrm{c-+iSWAP}$s, $4\left(\sqrt[3]{L_{c}}-\sqrt{L_{c}}\right)$
$\mathrm{c--iSWAP}$s , $8\left(L_{c}-\sqrt{L_{c}}\right)$ $\mathrm{c-}\mathbb{H}$
, $4\left(L_{c}-\sqrt{L_{c}}\right)$ $\mathrm{c-}\mathbb{J}$, $4\left(L_{c}-\sqrt{L_{c}}\right)$
$\mathrm{c-}\mathbb{J}^{\dagger}$, $4\left(L_{c}-\sqrt{L_{c}}\right)$
$\mathrm{c-}R_{\sigma_{x}}^{\Theta}$ and $4\left(L_{c}-\sqrt{L_{c}}\right)$
$\mathrm{c-}R_{\sigma_{y}}^{\Theta}$ are required to exactly implement
the evolution of $\mathcal{H_{\mathrm{kin}}}$. It may be possible
to reduce these numbers by some constant factor if the whole sequence
is precompiled and trivially cancelling operations are removed. The
alternance of the positive and negative variants of the $\mathrm{c-iSWAP}$
gates enforces the anticommutativity of the fermionic terms. The main
difficulties of the Fermi-Hubbard model arise from the fact that $\left[H_{\mathrm{kin}},H_{\mathrm{int}}\right]\neq0$,
a Trotter-Suzuki approximation must be used to evolve both terms at
the same time.

\subsubsection{S-wave pairing terms}

\begin{figure*}
\begin{centering}
\begin{minipage}[t]{1\columnwidth}%
\begin{center}
\subfigure[]{\label{fig:SimulationSPairingTerms}\Qcircuit @C=1em @R=.7em @!R { \lstick{P}					& \ctrl{1} & \ctrl{3} & \ctrl{5} & \ctrl{7} & \qw\\ \lstick{1\uparrow}	& \multigate{1}{R_{\Delta_s}^{\Theta_{\rm \Delta}}} & \qw & \qw & \qw & \qw \\ \lstick{1\downarrow}& \ghost{R_{\Delta_s}^{\Theta_{\rm \Delta}}} & \qw & \qw & \qw & \qw \\ \lstick{2\uparrow}	& \qw & \multigate{1}{R_{\Delta_s}^{\Theta_{\rm \Delta}}} & \qw & \qw & \qw \\ \lstick{2\downarrow}& \qw & \ghost{R_{\Delta_s}^{\Theta_{\rm \Delta}}} & \qw & \qw & \qw \\ \lstick{3\uparrow}	& \qw & \qw & \multigate{1}{R_{\Delta_s}^{\Theta_{\rm \Delta}}} & \qw & \qw \\ \lstick{3\downarrow}& \qw & \qw & \ghost{R_{\Delta_s}^{\Theta_{\rm \Delta}}} & \qw & \qw \\ \lstick{4\uparrow}	& \qw & \qw & \qw & \multigate{1}{R_{\Delta_s}^{\Theta_{\rm \Delta}}} & \qw \\ \lstick{4\downarrow}& \qw & \qw & \qw & \ghost{R_{\Delta_s}^{\Theta_{\rm \Delta}}} & \qw \\ & & & & &}}
\par\end{center}%
\end{minipage}
\par\end{centering}

\begin{centering}
\vfill{}
\begin{minipage}[t]{1\columnwidth}%
\begin{center}
\subfigure[]{\label{fig:SPairingRotationGate}\Qcircuit @C=1em @R=.7em @!R { \lstick{P} 						& \ctrl{1} 															& \qw & 																					& & \lstick{P}				 		& \ctrl{1} & \ctrl{1} & \ctrl{1} & \ctrl{1} & \ctrl{1} & \ctrl{1} & \ctrl{1} & \ctrl{1} & \ctrl{1} & \ctrl{1} & \qw \\ \lstick{i\uparrow} 		& \multigate{1}{R_{\Delta_s}^{\Theta}}	& \qw & \push{\rule{.3em}{0em}=\rule{.3em}{0em}}	& & \lstick{i\uparrow}		& \gate{\mathbb{H}} & \multigate{1}{-} & \gate{R_{\sigma_y}^{2\Theta}} & \multigate{1}{+} & \gate{\mathbb{H}} & \gate{\mathbb{J}} & \multigate{1}{-} & \gate{R_{\sigma_x}^{2\Theta}} & \multigate{1}{+} & \gate{\mathbb{J}^\dagger} & \qw\\ \lstick{i\downarrow}	& \ghost{R_{\Delta_s}^{\Theta}} 				& \qw & 																					& & \lstick{i\downarrow}	& \qw & \ghost{-} & \qw & \ghost{+} & \qw & \qw & \ghost{-} & \qw & \ghost{+} & \qw & \qw \\ & & & & & & & & & & & & & & & &}}
\par\end{center}%
\end{minipage}\caption{In \subref{fig:SimulationSPairingTerms}, the s-wave pairing terms
of the cluster Hamiltonian corresponding to $\mathcal{H}_{\mathrm{s-pair}}$
are decomposed into gates. The angle $\Theta_{{\rm \Delta}}\equiv-\Delta\tau\Delta_{s}'$.
There are $L_{c}$ terms like these in a square lattice. In \subref{fig:SPairingRotationGate}
, the decomposition of $\mathrm{c-}R_{\Delta_{s}}^{\Theta}$ in a
site subspace (spin $\uparrow$/$\downarrow$). There are $L_{c}$
terms like these in a square cluster. The single-qubit rotation gates
$R_{\sigma_{x}}^{\Theta}\equiv e^{-i\Theta\sigma_{x}}$ and $R_{\sigma_{y}}^{\Theta}\equiv e^{-i\Theta\sigma_{y}}$.
There are 6 $\mathrm{c-SQG}$s and 4 $\mathrm{c-\pm iSWAP}$s per
$\mathrm{c-}R_{\Delta_{s}}^{\Theta}$.\label{fig:SWave}}

\par\end{centering}

\end{figure*}
To verify that the ($U<0$) Fermi-Hubbard model supports s-wave superconductivity,
a variational singlet pairing term can be introduced as

\begin{equation}
\begin{array}{rcl}
\mathcal{H}_{\mathrm{s-pair}} & = & \Delta_{s}'\sum_{i}\left(c_{i\uparrow}^{\dagger}c_{i\downarrow}^{\dagger}+c_{i\downarrow}c_{i\uparrow}\right)\\
\\
 & = & \Delta_{s}'\sum_{i=1}^{L_{c}}\mathbb{D}_{L_{c}}\left(i\right),
\end{array}\label{eq:sPairingHamiltonian}
\end{equation}
where the $\mathbb{D}_{L_{c}}\left(i\right)$ strings are given by
\ref{eq:LocPairOperator}. From figures \ref{fig:SimulationSPairingTerms}
and \ref{fig:SPairingRotationGate}, it can seen that $2L_{c}$ $\mathrm{c-+iSWAP}$s,
$2L_{c}$ $\mathrm{c--iSWAP}$s , $2L_{c}$ $\mathrm{c-}\mathbb{H}$
, $L_{c}$ $\mathrm{c-}\mathbb{J}$, $L_{c}$ $\mathrm{c-}\mathbb{J}^{\dagger}$,
$L_{c}$ $\mathrm{c-}R_{\sigma_{x}}^{\Theta}$ and $L_{c}$ $\mathrm{c-}R_{\sigma_{y}}^{\Theta}$
are required to implement the evolution of $\mathcal{H}_{\mathrm{s-pair}}$.
The $\mathrm{c-SQG}$s are all operated on spin-$\uparrow$ orbitals.

\subsubsection{D-wave pairing terms}

\begin{figure*}
\begin{centering}
\begin{minipage}[t]{1\columnwidth}%
\begin{center}
\subfigure[]{\label{fig:SimulationDPairingTerms}\Qcircuit @C=1em @R=.7em @!R { \lstick{P}					& \ctrl{1} & \ctrl{1} & \ctrl{3} & \ctrl{5} & \ctrl{2} & \ctrl{2} & \ctrl{4} & \ctrl{6} & \qw\\ \lstick{1\uparrow}	& \multigate{3}{R_{\Delta_d\uparrow1}^{+\Theta_{\rm d}}} & \multigate{5}{R_{\Delta_d\uparrow2}^{+\Theta_{\rm d}}} & \qw & \qw & \qw & \qw & \qw & \qw & \qw \\ \lstick{1\downarrow}& \ghost{R_{\Delta_d\uparrow1}^{+\Theta_{\rm d}}} & \ghost{R_{\Delta_d\uparrow2}^{+\Theta_{\rm d}}} & \qw & \qw & \multigate{1}{R_{\Delta_d\downarrow1}^{-\Theta_{\rm d}}} & \multigate{3}{R_{\Delta_d\downarrow2}^{-\Theta_{\rm d}}} & \qw & \qw & \qw \\ \lstick{2\uparrow}	& \ghost{R_{\Delta_d\uparrow1}^{+\Theta_{\rm d}}} & \ghost{R_{\Delta_d\uparrow2}^{+\Theta_{\rm d}}} & \multigate{5}{R_{\Delta_d\uparrow2}^{+\Theta_{\rm d}}} & \qw & \ghost{R_{\Delta_d\downarrow1}^{-\Theta_{\rm d}}} & \ghost{R_{\Delta_d\downarrow2}^{-\Theta_{\rm d}}} & \qw & \qw & \qw \\ \lstick{2\downarrow}& \ghost{R_{\Delta_d\uparrow1}^{+\Theta_{\rm d}}} & \ghost{R_{\Delta_d\uparrow2}^{+\Theta_{\rm d}}} & \ghost{R_{\Delta_d\uparrow2}^{+\Theta_{\rm d}}} & \qw & \qw & \ghost{R_{\Delta_d\downarrow2}^{-\Theta_{\rm d}}} & \multigate{3}{R_{\Delta_d\downarrow2}^{-\Theta_{\rm d}}} & \qw & \qw \\ \lstick{3\uparrow}	& \qw & \ghost{R_{\Delta_d\uparrow2}^{+\Theta_{\rm d}}} & \ghost{R_{\Delta_d\uparrow2}^{+\Theta_{\rm d}}} & \multigate{3}{R_{\Delta_d\uparrow1}^{+\Theta_{\rm d}}} & \qw & \ghost{R_{\Delta_d\downarrow2}^{-\Theta_{\rm d}}} & \ghost{R_{\Delta_d\downarrow2}^{-\Theta_{\rm d}}} & \qw & \qw \\ \lstick{3\downarrow}& \qw & \ghost{R_{\Delta_d\uparrow2}^{+\Theta_{\rm d}}} & \ghost{R_{\Delta_d\uparrow2}^{+\Theta_{\rm d}}} & \ghost{R_{\Delta_d\uparrow1}^{+\Theta_{\rm d}}} & \qw & \qw & \ghost{R_{\Delta_d\downarrow2}^{-\Theta_{\rm d}}} & \multigate{1}{R_{\Delta_d\downarrow1}^{-\Theta_{\rm d}}} & \qw \\ \lstick{4\uparrow}	& \qw & \qw & \ghost{R_{\Delta_d\uparrow2}^{+\Theta_{\rm d}}} & \ghost{R_{\Delta_d\uparrow1}^{+\Theta_{\rm d}}} & \qw & \qw & \ghost{R_{\Delta_d\downarrow2}^{-\Theta_{\rm d}}} & \ghost{R_{\Delta_d\downarrow1}^{-\Theta_{\rm d}}} & \qw \\ \lstick{4\downarrow}& \qw & \qw & \ghost{R_{\Delta_d\uparrow2}^{+\Theta_{\rm d}}} & \ghost{R_{\Delta_d\uparrow1}^{+\Theta_{\rm d}}} & \qw & \qw & \qw & \qw & \qw \\ & & & & & & & & &}}
\par\end{center}%
\end{minipage}\vfill{}
\begin{minipage}[t]{1\columnwidth}%
\begin{center}
\subfigure[]{\label{fig:DPairingUpRotationGate}\Qcircuit @C=0.2em @R=.5em @!R { \lstick{P}								& \ctrl{1} 												& \qw & 																					& & \ctrl{1} & \ctrl{5} & \ctrl{4} & \push{\cdots} \qw & \ctrl{2} & \ctrl{1} & \ctrl{1} & \ctrl{1} & \ctrl{1} & \ctrl{1} & \ctrl{1} & \ctrl{1} & \ctrl{1} & \ctrl{2} & \push{\cdots} \qw & \ctrl{4} & \ctrl{5} & \ctrl{1} & \qw \\ \lstick{i\uparrow}				& \multigate{5}{R_{\Delta_d\uparrow m}^{\Theta}}	& \qw &																						& & \gate{\mathbb{H}} & \qw & \qw & \qw & \qw & \multigate{1}{-} & \gate{R_{\sigma_y}^{2\Theta}} & \multigate{1}{+} & \gate{\mathbb{H}} & \gate{\mathbb{J}} & \multigate{1}{-} & \gate{R_{\sigma_x}^{2\Theta}} & \multigate{1}{+} & \qw & \qw & \qw & \qw & \gate{\mathbb{J}^\dagger} & \qw \\ \lstick{i\downarrow}			& \ghost{R_{\Delta_d\uparrow m}^{\Theta}}					& \qw &																						& & \qw & \qw & \qw & \qw & \multigate{1}{+} & \ghost{-} & \qw & \ghost{+} & \qw & \qw & \ghost{-} & \qw & \ghost{+} & \multigate{1}{-} & \qw & \qw & \qw & \qw & \qw \\ \lstick{\vdots}						& \ghost{R_{\Delta_d\uparrow m}^{\Theta}}					& \qw & \push{\rule{.3em}{0em}=\rule{.3em}{0em}}	& & \qw & \qw & \qw & \push{\iddots}\qw & \ghost{+} & \qw & \qw & \qw & \qw & \qw & \qw & \qw & \qw & \ghost{-} & \push{\ddots}\qw & \qw & \qw & \qw & \qw \\ \lstick{\vdots}						& \ghost{R_{\Delta_d\uparrow m}^{\Theta}}					& \qw &																						& & \qw & \qw & \multigate{1}{+} & \push{\iddots}\qw & \qw & \qw & \qw & \qw & \qw & \qw & \qw & \qw & \qw & \qw  & \push{\ddots}\qw & \multigate{1}{-} & \qw & \qw & \qw \\ \lstick{(i+m)\uparrow}		& \ghost{R_{\Delta_d\uparrow m}^{\Theta}}					& \qw &																						& & \qw & \multigate{1}{-} & \ghost{+} & \qw & \qw & \qw & \qw & \qw & \qw & \qw & \qw & \qw & \qw & \qw & \qw & \ghost{-} & \multigate{1}{+} & \qw & \qw \\ \lstick{(i+m)\downarrow}	& \ghost{R_{\Delta_d\uparrow m}^{\Theta}}					& \qw &																						& & \qw & \ghost{-} & \qw & \qw & \qw & \qw & \qw & \qw & \qw & \qw & \qw & \qw & \qw & \qw & \qw & \qw & \ghost{+} & \qw & \qw \\ & & & & & & & & & & & & & & & & & & & & & & &}}
\par\end{center}%
\end{minipage}\vfill{}
\begin{minipage}[t]{1\columnwidth}%
\begin{center}
\subfigure[]{\label{fig:DPairingDownRotationGate}\Qcircuit @C=0.2em @R=.5em @!R { \lstick{P}								& \ctrl{1} 												& \qw & 																					& & \ctrl{1} & \ctrl{3} & \push{\cdots} \qw & \ctrl{1} & \ctrl{1} & \ctrl{1} & \ctrl{1} & \ctrl{1} & \ctrl{1} & \ctrl{1} & \ctrl{1} & \push{\cdots} \qw & \ctrl{3} & \ctrl{1} & \qw \\ \lstick{i\downarrow}				& \multigate{3}{R_{\Delta_d\downarrow m}^{\Theta}}	& \qw &																						& & \gate{\mathbb{H}} & \qw & \qw  & \multigate{1}{-} & \gate{R_{\sigma_y}^{2\Theta}} & \multigate{1}{+} & \gate{\mathbb{H}} & \gate{\mathbb{J}} & \multigate{1}{-} & \gate{R_{\sigma_x}^{2\Theta}} & \multigate{1}{+} & \qw & \qw & \gate{\mathbb{J}^\dagger} & \qw \\ \lstick{\vdots}			& \ghost{R_{\Delta_d\downarrow m}^{\Theta}}					& \qw &																						& & \qw & \qw & \push{\iddots}\qw & \ghost{-} & \qw & \ghost{+} & \qw & \qw & \ghost{-} & \qw & \ghost{+} & \push{\ddots}\qw & \qw & \qw & \qw \\ \lstick{\vdots}						& \ghost{R_{\Delta_d\downarrow m}^{\Theta}}					& \qw & \push{\rule{.3em}{0em}=\rule{.3em}{0em}}	& & \qw & \multigate{1}{-} & \push{\iddots}\qw & \qw & \qw & \qw & \qw & \qw & \qw & \qw & \qw & \push{\ddots}\qw & \multigate{1}{+} & \qw & \qw \\ \lstick{(i+m)\uparrow}						& \ghost{R_{\Delta_d\downarrow m}^{\Theta}}					& \qw &																						& & \qw & \ghost{-} & \qw & \qw & \qw & \qw & \qw & \qw & \qw & \qw & \qw & \qw & \ghost{+} & \qw & \qw \\ & & & & & & & & & & & & & & & & & & &}}
\par\end{center}%
\end{minipage}
\par\end{centering}

\centering{}\caption{In \subref{fig:SimulationDPairingTerms}, the d-wave pairing terms
of the cluster Hamiltonian corresponding to the time evolution of
$\mathcal{H}_{\mathrm{d_{x^{2}-y^{2}}}}$ are decomposed into gates.
The angle $\Theta_{{\rm d}}\equiv-\Delta\tau\Delta_{d}'$. There are
$4\left(L_{c}-\sqrt{L_{c}}\right)$ terms like these in a square lattice.
One quarter of those strings have length 2, another quarter has length
4, another quarter has length $2\sqrt{L_{c}}$ and the last quarter
has length $2\left(\sqrt{L_{c}}-1\right)$.  In \subref{fig:DPairingUpRotationGate},
the decomposition of $\mathrm{c-}R_{\Delta_{d}\uparrow m}^{\Theta}$
in a subspace starting at $i\uparrow$ and ending at $i+m\downarrow$
is shown, where $m=1$ or $\sqrt{L_{c}}$ in a square lattice. There
are 6 $\mathrm{c-SQG}$s and $4m+4$ $\mathrm{c-\pm iSWAP}$s per
$\mathrm{c-}R_{\Delta_{d}\uparrow m}^{\Theta}$. In \subref{fig:DPairingDownRotationGate},
the decomposition of $\mathrm{c-}R_{\Delta_{d}\downarrow m}^{\Theta}$
in a subspace starting at $i\downarrow$ and ending at $i+m\uparrow$
is shown, where $m=1$ or $\sqrt{L_{c}}$ in a square lattice. There
are 6 $\mathrm{c-SQG}$s and $4m$ $\mathrm{c-\pm iSWAP}$s per $\mathrm{c-}R_{\Delta_{d}\downarrow m}^{\Theta}$.\label{fig:DPairing}}
\end{figure*}
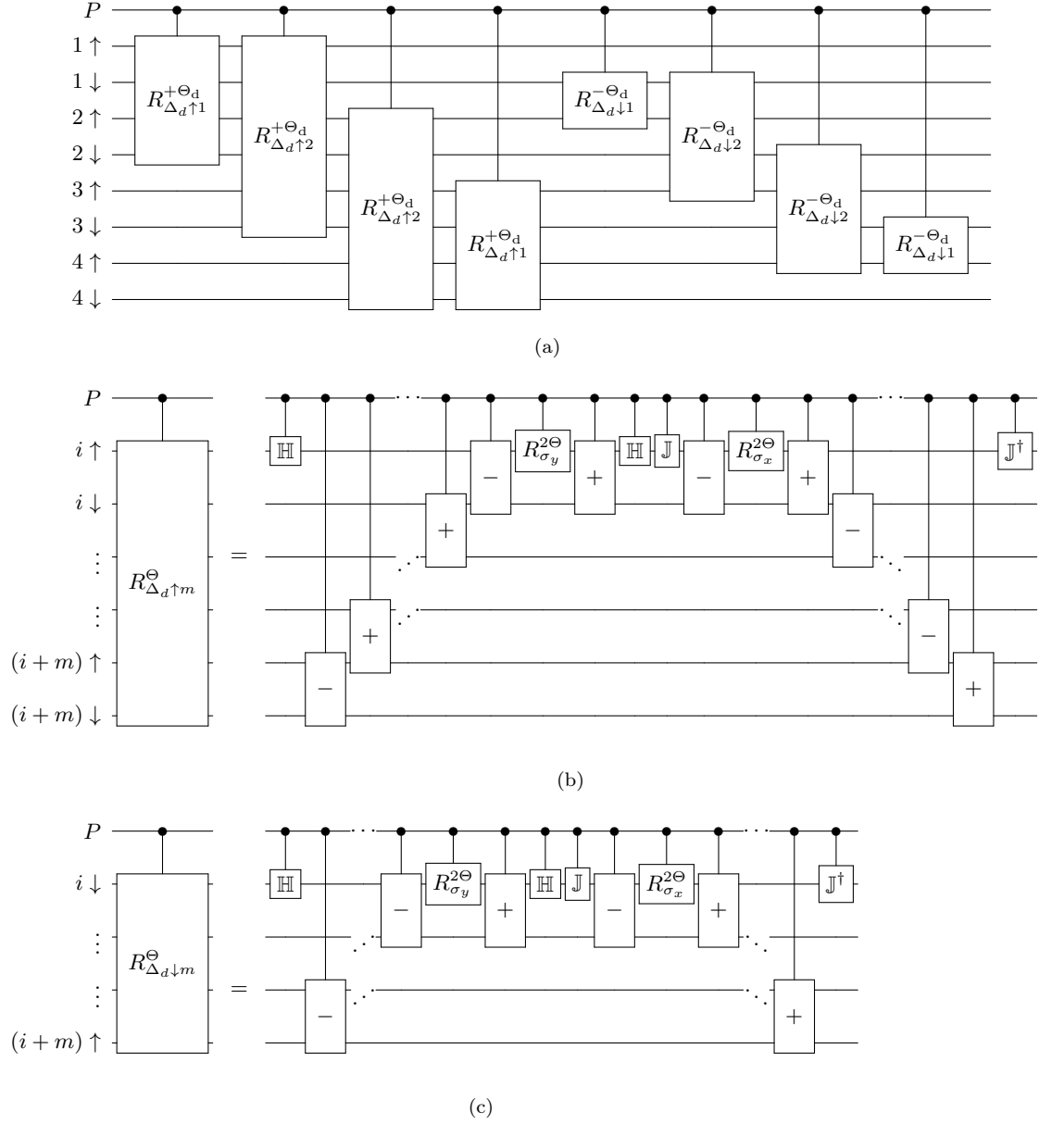
A superconducting $d_{x^{2}-y^{2}}$ singlet pairing term takes the
form \citep{Senechal08}
\begin{equation}
\begin{array}{rcl}
\mathcal{H}_{\mathrm{d_{x^{2}-y^{2}}}} & = & \Delta_{d}'\sum_{\left\langle i,j\right\rangle }\frac{d_{ij}}{2}\left(c_{i\uparrow}^{\dagger}c_{j\downarrow}^{\dagger}-c_{i\downarrow}^{\dagger}c_{j\uparrow}^{\dagger}+c_{j\downarrow}c_{i\uparrow}-c_{j\uparrow}c_{i\downarrow}\right)\\
\\
 & = & \Delta_{d}'\sum_{\left\langle i,j\right\rangle }\frac{d_{ij}}{2}\left(\mathbb{D}_{L_{c}\uparrow}\left(i,j\right)-\mathbb{D}_{L_{c}\downarrow}\left(i,j\right)\right)
\end{array}\label{eq:dParingHamiltonian}
\end{equation}
between nearest-neighbor site, where $\mathbf{R}$ are the vector
positions of the sites in the cluster in units of $a$ and
\begin{equation}
d_{ij}=\begin{cases}
1 & \mathrm{if}\:\mathbf{R}_{i}-\mathbf{R}_{j}=\pm a\mathbf{e_{x}}\\
-1 & \mathrm{if}\:\mathbf{R}_{i}-\mathbf{R}_{j}=\pm a\mathbf{e_{y}}\\
0 & \mathrm{otherwise.}
\end{cases}\label{eq:dwavePairingSymetry}
\end{equation}
The $\mathbb{D}_{L_{c}\sigma}\left(i,j\right)$ strings are given
by \ref{eq:PairOperator}. From figures \ref{fig:SimulationDPairingTerms},
\ref{fig:DPairingUpRotationGate} and \ref{fig:DPairingDownRotationGate},
it can seen that $4\left(\sqrt[3]{L_{c}}+L_{c}-2\sqrt{L_{c}}\right)$
$\mathrm{c-+iSWAP}$s, $4\left(\sqrt[3]{L_{c}}+L_{c}-\sqrt{L_{c}}\right)$
$\mathrm{c--iSWAP}$s , $8\left(L_{c}-\sqrt{L_{c}}\right)$ $\mathrm{c-}\mathbb{H}$
, $4\left(L_{c}-\sqrt{L_{c}}\right)$ $\mathrm{c-}\mathbb{J}$, $4\left(L_{c}-\sqrt{L_{c}}\right)$
$\mathrm{c-}\mathbb{J}^{\dagger}$, $4\left(L_{c}-\sqrt{L_{c}}\right)$
$\mathrm{c-}R_{\sigma_{x}}^{\Theta}$ and $4\left(L_{c}-\sqrt{L_{c}}\right)$
$\mathrm{c-}R_{\sigma_{y}}^{\Theta}$ are required to implement the
evolution of $\mathcal{H}_{\mathrm{d_{x^{2}-y^{2}}}}$. It may be
possible to reduce these numbers by some constant factor if the whole
sequence is precompiled and trivially cancelling operations are removed.
Interestingly, $\left[\mathcal{H_{\mathrm{kin}}},\mathcal{H}_{\mathrm{d_{x^{2}-y^{2}}}}\right]=0$
and the two terms of the cluster can be grouped together to simulate
their exact evolution.

\subsection{The Trotter-Suzuki approximation\label{sub:TrotterSuzuki}}

Typically, the terms of the cluster Hamiltonian (\ref{eq:HamiltonianOneCluster})
do not commute and a Trotter-Suzuki approximation \citep{Hatano05,Poulin15,LasHeras15}
must be used. Here is the procedure to make the mapping that requires
no oracle black box for $\mathcal{H}'$. The Hamiltonian (\ref{eq:HamiltonianOneCluster})
is broken into $M$ non-commuting parts such that

\begin{equation}
\mathcal{H}'=\sum_{i=1}^{M}\mathcal{H}_{i}'.\label{eq:TrotterHamiltonian}
\end{equation}
Each time-step $\Delta\tau$ evolution of the cluster Hamiltonian
can be simulated with $n_{T}$ Trotter-Suzuki steps

\begin{equation}
e^{-i\mathcal{H}'\Delta\tau}\simeq\left(\prod_{i=1}^{M}e^{-\frac{i\mathcal{H}_{i}'\Delta\tau}{n_{T}}}\right)^{n_{T}}+\sum_{i<j}\frac{\left[\mathcal{H}_{i}',\mathcal{H}_{j}'\right]\Delta\tau^{2}}{2n_{T}}+\ldots.\label{eq:TrotterFormula}
\end{equation}
It should be noted that those time-steps set the upper bound in the
simulated energy spectrum which should scale as $\omega_{\mathrm{max}}\propto\frac{1}{\Delta\tau}$,
while the lowest energy should scale at the inverse of the total simulation
time. 

The cluster Hamiltonian $\mathcal{H}'$ has 3 non-commuting blocks:
$\mathcal{H}_{z}\equiv\mathcal{H}_{\mathrm{local}}+\mathcal{H}_{\mathrm{int}}-\mathcal{H}_{AF}$,
$\mathcal{H_{\mathrm{kin}}}+\mathcal{H}_{\mathrm{d_{x^{2}-y^{2}}}}$
and $\mathcal{H}_{\mathrm{s-pair}}$, the commutation relations are
given in table (\ref{tab:CommutationRelations}). All blocks are skew-hermitians
such that $\mathcal{H}_{i}=\mathcal{H}_{i}^{*}$. The time evolution
of each time block can be done exactly. The blocks containing nearest-neighbor
operators ($\mathcal{H_{\mathrm{kin}}}$ and $\mathcal{H}_{\mathrm{d_{x^{2}-y^{2}}}}$)
are the most expensive in terms of gates. If $D$ is the dimension
of the lattice, then these blocks require the application of $O\left(L_{c}^{\frac{2D-1}{D}}\right)$
$\mathrm{c-\pm iSWAP}$s, so it is advisable to minimize the use of
these blocks in the Trotter-Suzuki decompostion. The number of gates
to implement the local interaction terms ($\mathcal{H}_{z}$ and $\mathcal{H}_{\mathrm{s-pair}}$)
scales as $O\left(L_{c}\right)$. 

\begin{table*}
\begin{centering}
\begin{tabular}{|c||c|c|c||c|c|c|}
\hline 
$\left[\bullet,\bullet\right]$ & $\mathcal{H}_{\mathrm{local}}$ & $\mathcal{H}_{\mathrm{int}}$ & $\mathcal{H}_{AF}$ & $\mathcal{H_{\mathrm{kin}}}$ & $\mathcal{H}_{\mathrm{s-pair}}$ & $\mathcal{H}_{\mathrm{d_{x^{2}-y^{2}}}}$\tabularnewline
\hline 
\hline 
$\mathcal{H}_{\mathrm{local}}$ & $0$ & $0$ & $0$ & $0$ & $-2\mathcal{H}_{D}$ & $-2\mathcal{H}_{F}$\tabularnewline
\hline 
$\mathcal{H}_{\mathrm{int}}$ & $0$ & $0$ & $0$ & $-2\mathcal{H}_{A}$ & $-\mathcal{H}_{E}$ & $-2\mathcal{H}_{G}$\tabularnewline
\hline 
$\mathcal{H}_{AF}$ & $0$ & $0$ & $0$ & $-2\mathcal{H}_{B}$ & $0$ & $-2\mathcal{H}_{H}$\tabularnewline
\hline 
\hline 
$\mathcal{H_{\mathrm{kin}}}$ & $0$ & $2\mathcal{H}_{A}$ & $2\mathcal{H}_{B}$ & $0$ & $\mathcal{H}_{C}$ & $0$\tabularnewline
\hline 
$\mathcal{H}_{\mathrm{s-pair}}$ & $2\mathcal{H}_{D}$ & $\mathcal{H}_{E}$ & $0$ & $-\mathcal{H}_{C}$ & $0$ & $0$\tabularnewline
\hline 
$\mathcal{H}_{\mathrm{d_{x^{2}-y^{2}}}}$ & $2\mathcal{H}_{F}$ & $2\mathcal{H}_{G}$ & $2\mathcal{H}_{H}$ & $0$ & $0$ & $0$\tabularnewline
\hline 
\end{tabular}
\par\end{centering}

\caption{Commutation relations of the different Hamiltonian terms ($2\times2$
cluster). $\mathcal{H}_{A}$ to $\mathcal{H}_{H}$ represent different
non-zero commutators.\label{tab:CommutationRelations}}

\end{table*}

\begin{figure}
\begin{centering}
\includegraphics[width=3.5in]{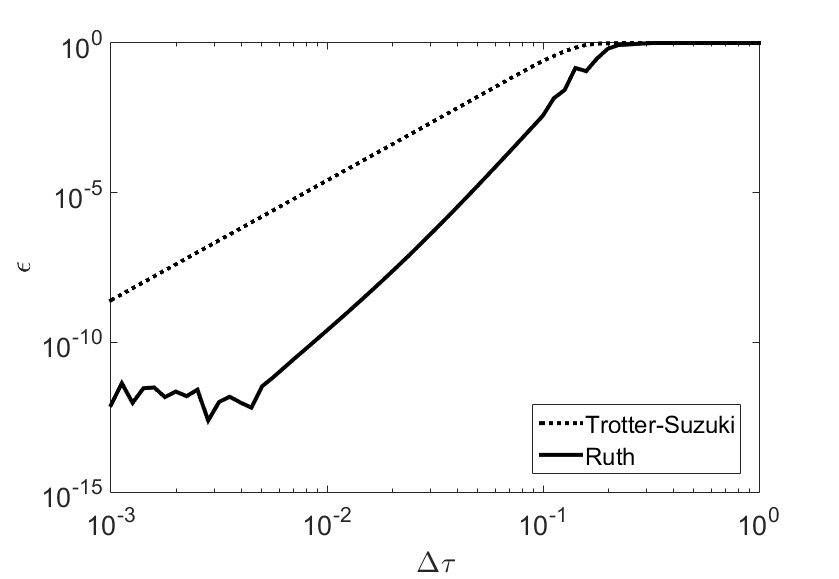}
\par\end{centering}

\caption{Numerical worst case error $\epsilon\left(\Delta\tau\right)=1-\frac{1}{16^{L_{c}}}\left|\mathrm{Tr}\left[U_{\mathrm{TS}}\left(N\Delta\tau\right)U^{\dagger}\left(N\Delta\tau\right)\right]\right|^{2}$for
the Trotter-Suzuki (in blue, order $O\left(\Delta\tau^{3}\right)$)
and the Ruth (in red, order $O\left(\Delta\tau^{4}\right)$) decompositions
for a constant simulation time such that $\tau=N\Delta\tau=3$. To
emulate a typical worst-case error, all variational parameters $\mu'=M'=\Delta_{s}'=\Delta_{d}'=3$.
The interaction $U=8$ and all energy and time units are made unitless
by referencing them to the hopping energy $t=1$. \label{fig:TrotterSuzukiError}}

\end{figure}

\begin{widetext}The worst-case Trotter-Suzuki decomposition arises
when all variational parameters have a non-zero value at some point
during the saddle-point search. In this case a single Trotter-Suzuki
step could be decomposed as
\begin{equation}
\ensuremath{\begin{array}{rcl}
e^{-i\mathcal{H}'\Delta\tau} & \approx & e^{-i\mathcal{H}_{z}\frac{\Delta\tau}{4}}\cdot e^{+i\mathcal{H}_{\mathrm{s-pair}}\frac{\Delta\tau}{2}}\cdot e^{-i\mathcal{H}_{z}\frac{\Delta\tau}{4}}\cdot e^{+i\mathcal{H}_{\mathrm{d_{x^{2}-y^{2}}}}\Delta\tau}\cdot e^{-i\mathcal{H_{\mathrm{kin}}}\Delta\tau}\ldots\\
 &  & \ldots\cdot e^{-i\mathcal{H}_{z}\frac{\Delta\tau}{4}}\cdot e^{+i\mathcal{H}_{\mathrm{s-pair}}\frac{\Delta\tau}{2}}\cdot e^{-i\mathcal{H}_{z}\frac{\Delta\tau}{4}}
\end{array}}\label{eq:TrotterSuzukiFull}
\end{equation}
Ruth's formula \citep{Ruth83,Hatano05} can also be used recursively

\begin{equation}
\ensuremath{e^{-i\Delta\tau\left(A+B\right)+O\left(\Delta\tau^{4}\right)}=e^{-i\frac{7}{24}\Delta\tau A}e^{-i\frac{2}{3}\Delta\tau B}e^{-i\frac{3}{4}\Delta\tau A}e^{+i\frac{2}{3}\Delta\tau B}e^{+i\frac{1}{24}\Delta\tau A}e^{-i\Delta\tau B}}\label{eq:RuthFormula}
\end{equation}
by replacing $A$ and $B$ by the correct cluster Hamiltonian terms.
\end{widetext}Ruth's formula is more precise but has a larger overhead
in term is gate count. In a Trotter-Suzuki step, the hopping term
$e^{-i\mathcal{H_{\mathrm{kin}}}\Delta\tau}$ and $e^{+i\mathcal{H}_{\mathrm{d_{x^{2}-y^{2}}}}\frac{\Delta\tau}{2}}$
appear once, the s-wave pairing term $e^{+i\mathcal{H}_{\mathrm{s-pair}}\frac{\Delta\tau}{4}}$
has two instances and the simple local $e^{-i\mathcal{H}_{z}\frac{\Delta\tau}{8}}$
appears four times. Figure \ref{fig:TrotterSuzukiError} provides
a practical effective bound on the error by looking at an extreme
case of non-commuting variational parameters all applied at the same
time. The error is given for a fixed evolution time by a varying step
size. A step size $\Delta\tau<10^{-2}$ achieve an error $\sim10^{-5}$
using a recursive Trotter-Suzuki formula and an error $\sim10^{-10}$
using a recursive Ruth formula. Not considering all variational parameters
at the same time significantly reduces the length of the decomposition. 

\section{Scaling to larger clusters\label{sec:Scaling}}

\begin{table}
\begin{centering}
\begin{tabular}{|c|c|c|c|c|c|c|c|c|}
\hline 
\begin{turn}{90}
Dimension(s)
\end{turn} & \begin{turn}{90}
Size
\end{turn} & \begin{turn}{90}
Orbitals (singlets) $\left[n\right]$
\end{turn} & \begin{turn}{90}
Dim. of Hilbert space $\left[2^{n}\right]$
\end{turn} & \begin{turn}{90}
Qubits required {[}$n+1${]}
\end{turn} & \begin{turn}{90}
Measured correl. functions$\left[<4n^{2}\right]$
\end{turn} & \begin{turn}{90}
$\mathrm{c-SQG}$s to tune $\left[7n\right]$
\end{turn} & \begin{turn}{90}
$\mathrm{c-\pm iSWAP}$s to tune $\left[2n-2\right]$
\end{turn} & \begin{turn}{90}
Gates / Trotter-Suzuki step (hopping terms)
\end{turn}\tabularnewline
\hline 
\hline 
1D & $2$ & $4$ & $16$ & $5$ & $64$ & $28$ & $6$ & $24$\tabularnewline
\hline 
1D & $3$ & $6$ & $64$ & $7$ & $144$ & $42$ & $10$ & $48$\tabularnewline
\hline 
1D & $4$ & $8$ & $256$ & $9$ & $256$ & $56$ & $14$ & $72$\tabularnewline
\hline 
2D & $2\times2$ & $8$ & $256$ & $9$ & $256$ & $56$ & $14$ & $96$\tabularnewline
\hline 
2D & $3\times3$ & $18$ & $262,144$ & $19$ & $1,296$ & $126$ & $34$ & $336$\tabularnewline
\hline 
2D & $4\times4$ & $32$ & $4,294,967,296$ & $33$ & $4,096$ & $224$ & $62$ & $768$\tabularnewline
\hline 
3D & $2\times2\times2$ & $16$ & $65,536$ & $17$ & $1,024$ & $112$ & $30$ & $416$\tabularnewline
\hline 
3D & $3\times3\times3$ & $54$ & $1.8\times10^{16}$ & $55$ & $11,664$ & $378$ & $106$ & $2,736$\tabularnewline
\hline 
3D & $4\times4\times4$ & $128$ & $3.4\times10^{38}$ & $129$ & $65,536$ & $896$ & $254$ & $10,368$\tabularnewline
\hline 
\end{tabular}
\par\end{centering}

\caption{Quantum ressources required to solve a cluster of the Fermi-Hubbard
once the Gibbs state is prepared. The information processed by the
classical computer is proportional to the number of measured correlation
functions which scales quadratically with the number of orbitals in
the cluster.\label{tab:QuantumRessourcesAdiabatic}}
\end{table}
The resource requirements of the algorithm are given in table \ref{tab:QuantumRessourcesAdiabatic}
by giving examples for the 1D, 2D and 3D Fermi-Hubbard model. The
1D model can be solved analytically and can be used as a benchmark.
The 3D model is meant to show that the method scales to higher dimensions.
All ressources only include the $P$ and $S$ registers, the scaling
of registers $R$ and $B$ are analyzed in details in \citep{Riera12}.While
the size of the Hilbert space required to store the density matrix
scales exponentially with the number of spin orbitals, the number
of qubits required in register $S$ scales linearly. The number of
correlation functions to measure, which corresponds to the amount
of classical information to extract from the quantum simulator, scales
quadratically with the size of the cluster.The number of conditional
single-qubits gates and the number of $\mathrm{c-\pm iSWAP}$s that
have to be benchmarked and tuned also scales linearly with the size
of the system, which is a significant technical advantage. Finally,
the number of $\mathrm{c-\pm iSWAP}$s in terms with nearest-neighbor
couplings (like hopping or d-wave superconductivity) scales subquadratically
as $O\left(L_{c}^{\frac{2D-1}{D}}\right)$, where $D$ is the dimension
of the system.

\section{Conclusion}

The Fermi-Hubbbard model contains the essential features of many strongly
correlated electronic systems. We recently proposed a method to compute
the properties of the Fermi-Hubbard using a hybrid quantum-classical
approach. In this paper we looked more closely at the scaling properties
of the quantum part of the algorithm by giving an explicit gate decomposition
of the time evolution of the cluster Hamiltonian and bounding expected
Trotter-Suzuki errors. The main results are the following:
\begin{enumerate}
\item It scales linearly in memory: 1 spin orbital corresponds to 1 qubit.
\item It scales favorably in number of measurements which are proportional
to $L_{c}^{2}$ at worst.
\item The number of time measurements determines precision in frequency
space (same as classical, decoherence means less information, ``good
enough'' is possible).
\item The most difficult terms require $O\left(L_{c}^{\frac{2D-1}{D}}\right)$
$\mathrm{c-\pm iSWAP}$s (the longest gate).
\item Trotter-Suzuki errors can be made as small as desired.
\item The proposed architecture has no crossing interaction lines whose
number scales as $O\left(L_{c}\right)$ with no long range interaction
required.
\item The number of gates that need to be tuned scales as $O\left(L_{c}\right)$.
\end{enumerate}
To fully benchmark the algorithm, a full simulation will have to be
implemented to analyze the gate count in the Gibbs state preparation.
A more careful analysis of errors also has to be done as the effect
of errors may not be the same depending if they appear in the $R$,
$P$ or $S+B$ registers. Finally, an adiabatic or annealing scheme
could be used to replace the Gibbs state preparation if only zero-temperature
states are studied \citep{Wecker15}. In this case, the correlation
function measurements would still stay the same as the rest of the
classical method.
\begin{acknowledgments}
This work was supported by SCALEQIT. The authors would like to thank
Ryan Babbush, David Poulin for very helpful discussions.
\end{acknowledgments}

\bibliographystyle{h-physrev}
\bibliography{Simulations}

\begin{thebibliography}{10}

\bibitem{Feynman82}
R.~P. Feynman,
\newblock International Journal of Theoretical Physics {\bf 21}, 467 (1982).

\bibitem{OMalley15}
P.~J.~J. O'Malley {\em et~al.},
\newblock Scalable quantum simulation of molecular energies, 2015.

\bibitem{Hubbard63}
J.~Hubbard,
\newblock Proceedings of the Royal Society of London. Series A, Mathematical
  and Physical Sciences {\bf 276}, 238 (1963).

\bibitem{Anderson87}
P.~W. Anderson,
\newblock Science {\bf 235}, 1196 (1987).

\bibitem{Kassal11}
I.~Kassal, J.~D. Whitfield, A.~Perdomo-Ortiz, M.-H. Yung, and A.~Aspuru-Guzik,
\newblock Annual Review of Physical Chemistry {\bf 62}, 185 (2011).

\bibitem{Peruzzo14}
A.~Peruzzo {\em et~al.},
\newblock Nature Communications {\bf 5}, 4213 (2014).

\bibitem{Babbush16}
R.~Babbush {\em et~al.},
\newblock New Journal of Physics {\bf 18}, 033032 (2016).

\bibitem{LasHeras13}
U.~L. Heras {\em et~al.},
\newblock Phys. Rev. Lett. {\bf 112}, 200501 (2013).

\bibitem{Lamata14}
L.~Lamata, A.~Mezzacapo, J.~Casanova, and E.~Solano,
\newblock EPJ Quantum Technology {\bf 1}, 9 (2014).

\bibitem{Barends15}
R.~Barends {\em et~al.},
\newblock Nature Communications {\bf 6}, 7654 (2015).

\bibitem{LasHeras15}
U.~L. Heras, L.~Garcia-Alvarez, A.~Mezzacapo, E.~Solano, and L.~Lamata,
\newblock EPJ Quantum Technology {\bf 2}, 8 (2015).

\bibitem{Bauer15}
B.~Bauer, D.~Wecker, A.~J. Millis, M.~B. Hastings, and M.~Troyer,
\newblock Hybrid quantum-classical approach to correlated materials, 2015.

\bibitem{Salathe15}
Y.~Salathe {\em et~al.},
\newblock Phys. Rev. X {\bf 5}, 021027 (2015).

\bibitem{Potthoff03}
M.~Potthoff, M.~Aichhorn, and C.~Dahnken,
\newblock Phys. Rev. Lett. {\bf 91}, 206402 (2003).

\bibitem{DDW16}
P.-L. Dallaire-Demers and F.~K. Wilhelm,
\newblock Phys. Rev. A {\bf 93}, 032303 (2016).

\bibitem{Thompson13}
J.~Thompson, M.~Gu, K.~Modi, and V.~Vedral,
\newblock Quantum computing with black-box subroutines, 2013.

\bibitem{Lieb03}
E.~H. Lieb and F.~Y. Wu,
\newblock Physica A {\bf 321}, 1 (2003).

\bibitem{Senechal05}
D.~Sénéchal, P.-L. Lavertu, M.-A. Marois, and A.~Tremblay,
\newblock Phys. Rev. Lett. {\bf 94}, 156404 (2005).

\bibitem{Tremblay06}
A.-M. Tremblay, B.~Kyung, and D.~Sénéchal,
\newblock Low Temperature Physics {\bf 32}, 424 (2006).

\bibitem{Senechal08}
D.~Senechal,
\newblock An introduction to quantum cluster methods, 2008,
  cond-mat.str-el/0806.2690v2.

\bibitem{Riera12}
A.~Riera, C.~Gogolin, and J.~Eisert,
\newblock Phys. Rev. Lett. {\bf 108}, 080402 (2012).

\bibitem{Kaicher16}
M.~Kaicher, F.~Motzoi, and F.~K. Wilhelm,
\newblock Pauli strings with the exchange interaction,
\newblock 2016.

\bibitem{Poulin15}
D.~Poulin {\em et~al.},
\newblock QIC {\bf 15}, 361 (2015).

\bibitem{Jordan28}
P.~Jordan and E.~Wigner,
\newblock Z. Phys. {\bf 47}, 631 (1928).

\bibitem{Nielsen01}
M.~A. Nielsen and I.~L. Chuang,
\newblock {\em Quantum Computation and Quantum Information} (Cambridge
  University Press, 2001).

\bibitem{Liebermann16}
P.~Liebermann, P.-L. Dallaire-Demers, E.~Assémat, and F.~K. Wilhelm,
\newblock Conditional imaginary swap gates for quantum simulation of fermions,
\newblock 2016.

\bibitem{Hatano05}
N.~Hatano and M.~Suzuki,
\newblock {\em Quantum Annealing and Other Optimization Methods}, Lecture Notes
  in Physics Vol. 679 (Springer Berlin Heidelberg, 2005), chap. Finding
  Exponential Product Formulas of Higher Orders, pp. 37--68.

\bibitem{Ruth83}
R.~D. Ruth,
\newblock IEEE Transactions on Nuclear Science {\bf 30}, 2669 (1983).

\bibitem{Wecker15}
D.~Wecker {\em et~al.},
\newblock Phys. Rev. A {\bf 92}, 062318 (2015).

\end{thebibliography}

\end{document}